\documentclass[iop]{emulateapj}
\usepackage{graphicx}
\usepackage{amsmath}
\usepackage{rotating}
\usepackage{threeparttable}
\usepackage{multirow}

\shorttitle{Variability-Selected LLAGNs in the 7 Ms \mbox{CDF-S}}

\begin{document}

\title{VARIABILITY-SELECTED LOW-LUMINOSITY ACTIVE GALACTIC NUCLEI CANDIDATES IN THE 7 MS CHANDRA DEEP FIELD-SOUTH}

\author{\mbox{N. Ding}\altaffilmark{1,2,3}, \mbox{B. Luo}\altaffilmark{1,2,3}, \mbox{W. N. Brandt}\altaffilmark{4,5,6}, \mbox{M. Paolillo}\altaffilmark{7,8,9}, \mbox{G. Yang}\altaffilmark{4,5}, \mbox{B. D. Lehmer}\altaffilmark{10}, \mbox{O. Shemmer}\altaffilmark{11}, \mbox{D. P. Schneider}\altaffilmark{4, 5}, \mbox{P. Tozzi}\altaffilmark{12}, \mbox{Y. Q. Xue}\altaffilmark{13, 14}, \mbox{X. C. Zheng}\altaffilmark{13, 14}, \mbox{Q. S. Gu}\altaffilmark{1,2,3}, \mbox{A. M. Koekemoer}\altaffilmark{15}, \mbox{C. Vignali}\altaffilmark{16,17}, \mbox{F. Vito}\altaffilmark{4,5}, \mbox{J. X. Wang}\altaffilmark{13, 14}}

\altaffiltext{1}{School of Astronomy and Space Science, Nanjing University, Nanjing, Jiangsu 210093, China}
\altaffiltext{2}{Key Laboratory of Modern Astronomy and Astrophysics (Nanjing University), Ministry of Education, Nanjing 210093, China}
\altaffiltext{3}{Collaborative Innovation Center of Modern Astronomy and Space Exploration, Nanjing 210093, China}
\altaffiltext{4}{Department of Astronomy and Astrophysics, 525 Davey Lab, The Pennsylvania State University, University Park, PA 16802, USA}
\altaffiltext{5}{Institute for Gravitation and the Cosmos, The Pennsylvania State University, University Park, PA 16802, USA}
\altaffiltext{6}{Department of Physics, 104 Davey Lab, The Pennsylvania State University, University Park, PA 16802, USA}
\altaffiltext{7}{Dipartimento di Fisica “Ettore Pancini”, Universit\`{a} di Napoli Federico II, via Cintia, 80126, Italy}
\altaffiltext{8}{INAF - Osservatorio Astronomico di Capodimonte, via Moiariello 16, 80131, Napoli, Italy}
\altaffiltext{9}{INFN - Unit\`{a} di Napoli, via Cintia 9, 80126, Napoli, Italy}
\altaffiltext{10}{Department of Physics, University of Arkansas, 226 Physics Building, 825 West Dickson Street, Fayetteville, AR 72701, USA}
\altaffiltext{11}{Department of Physics, University of North Texas, Denton, TX 76203, USA}
\altaffiltext{12}{INAF -- Osservatorio Astrofisico di Firenze, Largo Enrico Fermi 5, I-50125 Firenze, Italy}
\altaffiltext{13}{CAS Key Laboratory for Research in Galaxies and Cosmology, Department of Astronomy, University of Science and Technology of China, Hefei 230026, China}
\altaffiltext{14}{School of Astronomy and Space Science, University of Science and Technology of China, Hefei 230026, China}
\altaffiltext{15}{Space Telescope Science Institute, 3700 San Martin Drive, Baltimore MD 21218, USA}
\altaffiltext{16}{Dipartimento di Fisica e Astronomia, Universit\`{a} degli Studi di Bologna, via Gobetti 93/2, 40129 Bologna, Italy}
\altaffiltext{17}{INAF -- Osservatorio Astronomico di Bologna, via Gobetti 93/3, 40129 Bologna, Italy}

\begin{abstract}
In deep X-ray surveys, active galactic nuclei (AGNs) with a broad range of luminosities have been identified. However, cosmologically distant low-luminosity AGN (LLAGN, $L_{\mathrm{X}} \lesssim 10^{42}$~erg~s$^{-1}$) identification still poses a challenge due to significant contamination from host galaxies. Based on the 7~Ms \emph{Chandra} Deep Field-South (\mbox{CDF-S}) survey, the longest timescale ($\sim 17$ years) deep \mbox{X-ray} survey to date, we utilize an \mbox{X-ray} variability selection technique to search for LLAGNs that remain unidentified among the \mbox{CDF-S} \mbox{X-ray} sources. We find 13 variable sources from 110 unclassified \mbox{CDF-S} \mbox{X-ray} sources. Except for one source which could be an ultraluminous \mbox{X-ray} source, the variability of the remaining 12 sources is most likely due to accreting supermassive black holes. These 12 AGN candidates have low intrinsic \mbox{X-ray} luminosities, with a median value of \mbox{$7 \times10^{40}$~erg~s$^{-1}$}. They are generally not heavily obscured, with an average effective power-law photon index of 1.8. The fraction of variable AGNs in the \mbox{CDF-S} is independent of \mbox{X-ray} luminosity and is only restricted by the total number of observed net counts, confirming previous findings that \mbox{X-ray} variability is a \mbox{near-ubiquitous} property of AGNs over a wide range of luminosities. There is an anti-correlation between \mbox{X-ray} luminosity and variability amplitude for high-luminosity AGNs, but as the luminosity drops to $\lesssim 10^{42}$~erg~s$^{-1}$, the variability amplitude no longer appears dependent on the luminosity. The entire observed \mbox{luminosity-variability} trend can be roughly reproduced by an empirical AGN variability model based on a broken power-law power spectral density function.\\
\\
\end{abstract}


\section{INTRODUCTION}
It is generally believed that a supermassive black hole (SMBH) is located in the center of an active galactic nucleus (AGN). The SMBH provides energy for the AGN by accreting the surrounding material. 
With a range of X-ray surveys from shallow all-sky surveys to \hbox{ultradeep} pencil-beam surveys, we can obtain a relatively complete census of AGN populations over a wide redshift range \citep[e.g.,][]{2015A&ARv..23....1B, 2017arXiv170904601X}. These sensitive \mbox{X-ray} surveys provide an unprecedented opportunity to trace the growth process of SMBHs and investigate the evolution of the AGN luminosity function \citep[e.g.,][]{2017MNRAS.471.4398P, 2017arXiv170907892V}.

In deep \mbox{X-ray} surveys, a source is identified as an AGN according to a number of empirical criteria, for example, a high intrinsic rest-frame \mbox{X-ray} luminosity ($L_{\mathrm{X}} >3 \times 10^{42}~\mathrm{erg~s}^{-1}$), a hard \mbox{X-ray} spectrum (effective power-law spectral photon index less than one), or a large ratio of \mbox{X-ray} flux to \emph{R}-band flux ($\log  F_{X}/F_{R} > -1$) (see \citealt[][]{2011ApJS..195...10X,2016ApJS..224...15X,2017ApJS..228....2L} for details). These selection criteria are successful in selecting AGNs with a broad range of luminosities, but they nevertheless can miss certain populations, such as heavily obscured AGNs and \mbox{low-luminosity} AGNs (see, e.g., \citealt{2012ApJ...748..124Y, 2015A&ARv..23....1B, 2017A&ARv..25....2P}). \mbox{Low-luminosity} AGNs (LLAGNs) have low \mbox{X-ray} luminosities ($L_{\mathrm{X}} \lesssim 10^{42}~\mathrm{erg~s}^{-1}$) and often low Eddington accretion rates ($\lambda_{\mathrm{Edd}} \ll 0.1$) relative to ordinary AGNs (see, e.g., \citealt[][]{2008ARA&A..46..475H,2017A&ARv..25....2P}). Despite the lack of violent nuclear activities, they are still a very important AGN population for understanding the transition of accretion modes, the growth of SMBHs, and the coevolution of AGNs and host galaxies (see \citealt[][]{2008ARA&A..46..475H} for a review). Therefore, it is important to improve the completeness of the census of LLAGNs in deep \mbox{X-ray} surveys. \citet[][]{2006AJ....131..133P} simulated \emph{Chandra} observations of nearby low-luminosity Seyfert nuclei artificially shifted to redshift $\sim0.3$. Their results suggested that these LLAGNs will exhibit the same \mbox{X-ray} luminosities, spectral shapes and X-ray-to-optical flux ratios as those of normal or optically bright, X-ray-faint galaxies, which indicates that some cosmologically distant LLAGNs still cannot be identified relying on the above characteristics (i.e., \mbox{X-ray} luminosity, spectral shape, etc). Cosmologically distant LLAGN identification poses a great challenge.

Variability of the multiwavelength (from radio to \hbox{$\gamma$-ray}) radiation is one of the distinctive characteristics of AGNs. Therefore, variability can be used as a tool to select AGNs in extragalactic surveys. For example, \citet{2008A&A...488...73T}, \citet{2008ApJ...676..121M}, \citet{2010ApJ...723..737V}, \citet{2011ApJ...731...97S}, and \citet[][2018 in prep]{2016ASSP...42..269D} used optical variability to select AGNs from the 1~Ms \emph{Chandra} Deep \mbox{Field-South} (\mbox{CDF-S}), the Subaru/\emph{XMM-Newton} Deep Field, the \mbox{GOODS-North} Field, the \mbox{GOODS-South} Field, and the \mbox{VST-SUDARE} survey of the COSMOS Field, respectively.
Because AGNs are more variable in the \mbox{X-ray} band than in the UV/optical band, and AGN emission in the \hbox{X-ray} band has relatively less contamination from \mbox{non-AGN} systems, \hbox{X-ray} variability can be effectively used as a tool to identify LLAGNs \citep[e.g.,][]{2012ApJ...748..124Y}. The variability of AGNs generally presents ``red noise'' behavior, namely the occurrence of larger amplitude variations on longer timescales \citep[e.g.,][]{1987Natur.325..694L,2006Natur.444..730M}. As such, AGN variability is more easily detected in a longer observational time span \citep[see, e.g.,][]{2014ApJ...781..105L,2017A&ARv..25....2P}. \citet{2012ApJ...748..124Y} utilized the \mbox{X-ray} variability selection technique to search for AGN candidates missed by other selection criteria in the 4~Ms \mbox{CDF-S} \citep{2011ApJS..195...10X}. They reported that 20 \mbox{CDF-S} galaxies have \hbox{long-term} \hbox{X-ray} variability, and 19 of them are LLAGNs.

Recently, the 7~Ms \mbox{CDF-S} survey, the deepest \mbox{X-ray} survey to date, was presented in \citet{2017ApJS..228....2L}. This survey dataset provides a lengthy observational time span (\mbox{$\sim 17$~years}) and a large number of observed net counts for the \mbox{X-ray} sources, improving our ability to detect \mbox{X-ray} variability. Therefore, we use the 7~Ms \mbox{CDF-S} data to search for LLAGNs from unclassified \mbox{X-ray} sources in the \mbox{CDF-S}.
This paper is organized as follows. Section~2 describes data processing and sample selection. Section~3 presents the method and results of searching for variable \mbox{X-ray} sources. We further select LLAGN candidates from these variable \mbox{X-ray} sources and report their basic properties in Section~4. In Section~5 we discuss the efficiency of \mbox{X-ray} variability selection of LLAGNs and the fraction of variable AGNs in the \mbox{CDF-S}, and also explore the relation between \mbox{X-ray} luminosity and variability amplitude for \mbox{long-term} AGN variability. Section~6 summarizes the main results of this work. Cosmological parameters of $H_{0}=70~$km~s$^{-1}$Mpc$^{-1}$, $\Omega_{m}=0.3$, and $\Omega_{\Lambda}=0.7$ \citep[][]{2016A&A...594A..13P} and a Chabrier initial mass function (IMF; \citealt{2003ApJ...586L.133C}) are adopted in this work.

\section{DATA AND SAMPLE}
This study is based on the 7~Ms CDF-S data \citep{2017ApJS..228....2L}. The dataset consists of 102 observations collected by the \emph{Chandra} Advanced CCD Imaging Spectrometer imaging array (ACIS-I; \citealt{2003SPIE.4851...28G}) from October 1999 to March 2016, and the cumulative observed exposure time is nearly 7~Ms. We divide the CDF-S observations into seven epochs according to the distribution of the observation times (see Table~\ref{Table1} for details); the first four epochs are the same as those in \citet[][]{2012ApJ...748..124Y}. Each epoch spans several months to one year and has approximately 1~Ms combined exposure time. This binning strategy enhances the \hbox{signal-to-noise} ratio for faint sources while minimizing the influence of broad time-span bins on variability.

The 7~Ms CDF-S main catalog contains 1008 \mbox{X-ray} sources, of which 711 are classified as AGNs. Excluding 12 stars, the remaining 285 unclassified \mbox{X-ray} sources are considered as normal galaxies (see \citealt{2017ApJS..228....2L} for details). To perform reliable variability analyses, we construct a sample based on the following four criteria:

1. The source has an off-axis angle of $<8'$;

2. The source has at least 30 net 0.5--7~keV counts in the full 7~Ms exposure;

3. The source is not one of the six transient events identified by \citet{2017arXiv171004358Z}.

4. The source has redshift measurements.

The first criterion ensures that each source is well covered by all observations. The second criterion excludes sources with extremely weak \hbox{X-ray} emission to improve the reliability of variability analyses. The third criterion excludes transient events, because they are probably not bona fide AGNs. There are five sources that satisfy the first two criteria but do not meet the third criterion. Their XIDs\footnote{XID in the entire paper refers to the source sequence number in the 7~Ms main catalog.} are 297, 330, 403, 725, 935. They have all been classified as AGNs in \citet{2017ApJS..228....2L}. This conservative criterion (i.e., excluding the five controversial AGNs from the AGN sample) does not materially affect the results of our following analyses. The fourth criterion enables us to calculate \mbox{X-ray} luminosity for each source. 
There are six sources that are excluded from the sample due to this criterion. 
The six sources have all been classified as AGNs in \citet{2017ApJS..228....2L}, and they are unlikely LLAGNs at low redshifts. We obtain a final sample containing 505 sources, of which 395 are AGNs and 110 are unclassified \mbox{X-ray} sources. These unclassified \mbox{X-ray} sources are considered normal galaxies in \citet{2017ApJS..228....2L} but may actually contain unidentified LLAGNs, and they are the main subject investigated in this paper. We examine the variability of the 395 AGNs for comparison purposes in Section~5. We use ACIS Extract \citep[AE,][]{2010ApJ...714.1582B}\footnote{See http://www.astro.psu.edu/xray/docs/TARA/ae\_users\_guide.html for details on ACIS Extract.} to extract source photometry in each epoch. The extracted full-band (0.5--7~keV) photometry is utilized to analyze variability.\footnote{We only use the \mbox{full-band} data (instead of \mbox{soft-band} or \mbox{hard-band} data) to analyze variability, because the largest number of net counts in the full band for a given source improves the likelihood and reliability of variability detection.}

\begin{table*}
\tablenum{1}
\centering
\caption{Details of the Seven Epochs}
\scriptsize
\begin{tabular}{ccc}
\hline\hline
\multirow{2}{*}{Epoch} & Observation Date Range and &	\multirow{2}{*}{Total Exposure Time (ks)}\\
&Observation IDs&\\
\hline
\multirow{2}{*}{1} &1999.10 -- 2000.12 & \multirow{2}{*}{930.8} \\
& 1431-0 1431-1 441   582  2406  2405  2312  1672  2409  2313  2239	& \\
\hline
\multirow{2}{*}{2} &2007.09 -- 2007.11 & \multirow{2}{*}{959.9} \\
& 8591  9593  9718  8593  8597  8595  8592  8596  9575  9578  8594  9596	& \\
\hline
\multirow{2}{*}{3} &2010.03 -- 2010.05 & \multirow{2}{*}{1012.6} \\
& 12043 12123 12044 12128 12045 12129 12135 12046 12047 12137 12138 12055 12213 12048& \\
\hline
\multirow{2}{*}{4} &2010.05 -- 2010.07 & \multirow{2}{*}{943.3} \\
& 12049 12050 12222 12219 12051 12218 12223 12052 12220 12053 12054 12230 12231 12227 12233 12232 12234& \\
\hline
\multirow{2}{*}{5} &2014.06 -- 2014.10 & \multirow{2}{*}{931.4} \\
& 16183 16180 16456 16641 16457 16644 16463 17417 17416 16454 16176 16175 16178 16177 16620 16462& \\
\hline
\multirow{2}{*}{6} &2014.10 -- 2015.01& \multirow{2}{*}{922.2} \\
& 17535 17542 16184 16182 16181 17546 16186 16187 16188 16450 16190 16189 17556 16179 17573& \\
\hline
\multirow{2}{*}{7} &2015.03 -- 2016.03& \multirow{2}{*}{1026.8} \\
& 17633 17634 16453 16451 16461 16191 16460 16459 17552 16455 16458 17677 18709 18719 16452 18730 16185& \\
\hline
\label{Table1}
\end{tabular}
\end{table*}

\section{Searching for variable \mbox{X-ray} sources}
In this section, we search for variable \mbox{X-ray} sources from the 110 unclassified X-ray sources. These variable \mbox{X-ray} sources will be investigated further in the next section (Section~4) regarding whether their \hbox{X-ray} variability is due to accreting SMBHs.

\subsection{Method}
We assess whether a source has \hbox{X-ray} variability according to the probability that its observed \hbox{X-ray} photon flux fluctuation is generated by Poisson noise alone. \citep[e.g.,][]{2012ApJ...748..124Y, 2016ApJ...831..145Y, 2017MNRAS.471.4398P}. 

We utilize the statistic
\begin{equation}
\label{eq1}
X^2 = \sum_{i=1}^7\frac{(PF_{i} - \langle PF \rangle)^2}{(\delta PF_{i})^2}
\end{equation}
to measure \hbox{X-ray} photon flux fluctuation, where $i$ indicates different epochs (1--7); $PF_{i}$ is the photon flux in each epoch; $\langle PF \rangle$ is the measured photon flux from the stacked 7~Ms data. The photon flux in each epoch is calculated as
\begin{equation}
\label{eq2}
PF_{i} = \frac{NET\_CNTS_{i}}{EFFAREA_{i} \times EXPOSURE_{i} \times PSF\_FRAC_{i}},
\end{equation}
and its 1$\sigma$ uncertainty is
\begin{equation}
\label{eq3}
\delta PF_{i} = \frac{\delta NET\_CNTS_{i}}{EFFAREA_{i} \times EXPOSURE_{i} \times PSF\_FRAC_{i}},
\end{equation}
where $NET\_CNTS_{i}$ is the background-subtracted counts (i.e., net counts); $\delta NET\_CNTS_{i}$ is the uncertainty on the net counts, which is the mean of the upper and lower 1$\sigma$ errors;\footnote{The upper and lower 1$\sigma$ errors are calculated following \citet{1986ApJ...303..336G}.} the $EFFAREA_{i}$, $EXPOSURE_{i}$, and $PSF\_FRAC_{i}$ parameters are the effective area, exposure time, and PSF fraction calculated by AE, respectively. The effective area takes into account the varying sensitivity of \emph{Chandra} observations between the different epochs (e.g., due to vignetting, CCD gaps, quantum-efficiency degradation).

For a non-variable source, its observed photon flux fluctuation is entirely due to Poisson noise. When the count rate of such a source is high ($\gtrsim 15$ net counts in each epoch; e.g., \citealt{2012ApJ...748..124Y}), the $X^2$ value of its photon flux fluctuation will follow a $\chi^2$ distribution with six degrees of freedom. However, when the count rate is low ($\lesssim 15$ net counts in each epoch), the $X^2$ value will deviate from the $\chi^2$ distribution because the errors on the net counts are not Gaussian \citep[e.g.,][]{2004ApJ...611...93P}. In our sample, most sources have low counts in each epoch bin. Therefore, we use Monte Carlo simulations to obtain the $X^2$ value of photon flux fluctuation produced by Poisson noise alone \citep[e.g.,][]{2004ApJ...611...93P, 2012ApJ...748..124Y, 2016ApJ...831..145Y, 2017MNRAS.471.4398P}. By comparing the measured $X^2$ value to the $X^2$ values generated from the simulations, we can obtain the probability that the observed photon flux fluctuation is produced by Poisson noise alone. The specific process is described as follows:

In each simulation, we assume that the source is non-variable and $PF$ remains constant in all seven epochs. The exact value of $PF$ should be the measured $\langle PF \rangle$ value modified by its statistical uncertainty. We test to consider the effects of uncertainty on $\langle PF \rangle$ and find that these have negligible effect on the simulation result. Therefore, we simply adopt the measured $\langle PF \rangle$ as $PF$ in each simulation. Once the $PF$ value is determined, the model net counts, model background counts, and model source counts are calculated as

\begin{equation}
\label{eq4}
\begin{split}
NET\_CNTS_{i}^\mathrm{model} = \langle PF \rangle \times EFFAREA_{i} \\
\times EXPOSURE_{i} \times PSF\_FRAC_{i},
\end{split}
\end{equation}
\begin{equation}
\label{eq5}
BKG\_CNTS_{i}^\mathrm{model} = BKG\_CNTS_{i}^\mathrm{observed},
\end{equation}
\begin{equation}
\label{eq6}
\begin{split}
SRC\_CNTS_{i}^\mathrm{model} = NET\_CNTS_{i}^\mathrm{model} \\
+ \frac{BKG\_CNTS_{i}^\mathrm{model}}{BACKSCAL_{i}},
\end{split}
\end{equation}
where $BACKSCAL_{i}$ is the area ratio of the \mbox{background-extraction} aperture to the \mbox{source-extraction} aperture. We simulate the observed source counts (background counts) by randomly sampling from a Poisson distribution with a mean value of $SRC\_CNTS_{i}^\mathrm{model} $ ($BKG\_CNTS_{i}^\mathrm{model}$). We then extract source photometry (net counts and errors) following the algorithm in AE and calculate the $X^2$ value of the simulated data using \hbox{Eqs. 1--3}.
The simulations are performed 10000 times to obtain 10000 $X^2$ values. By comparing the distribution of these simulated $X^2$ values to the $X^2$ value of the observed photon flux fluctuation, we obtain the proportion of the simulated $X^2$ values greater than the measured $X^2$ value. This proportion represents the probability ($P_{X^2}$) that the observed photon flux fluctuation is generated by Poisson noise alone. For a given source, if none of its 10000 simulated $X^2$ values is larger than its measured $X^2$ value, we obtain a $90\%$ confidence level upper limit on its $P_{X^2}$ value, i.e., $P_{X^2}<0.0002$ \citep{1986ApJ...303..336G}.

\subsection{Results}
\begin{figure*}[]
\centering
\includegraphics[width=7.in]{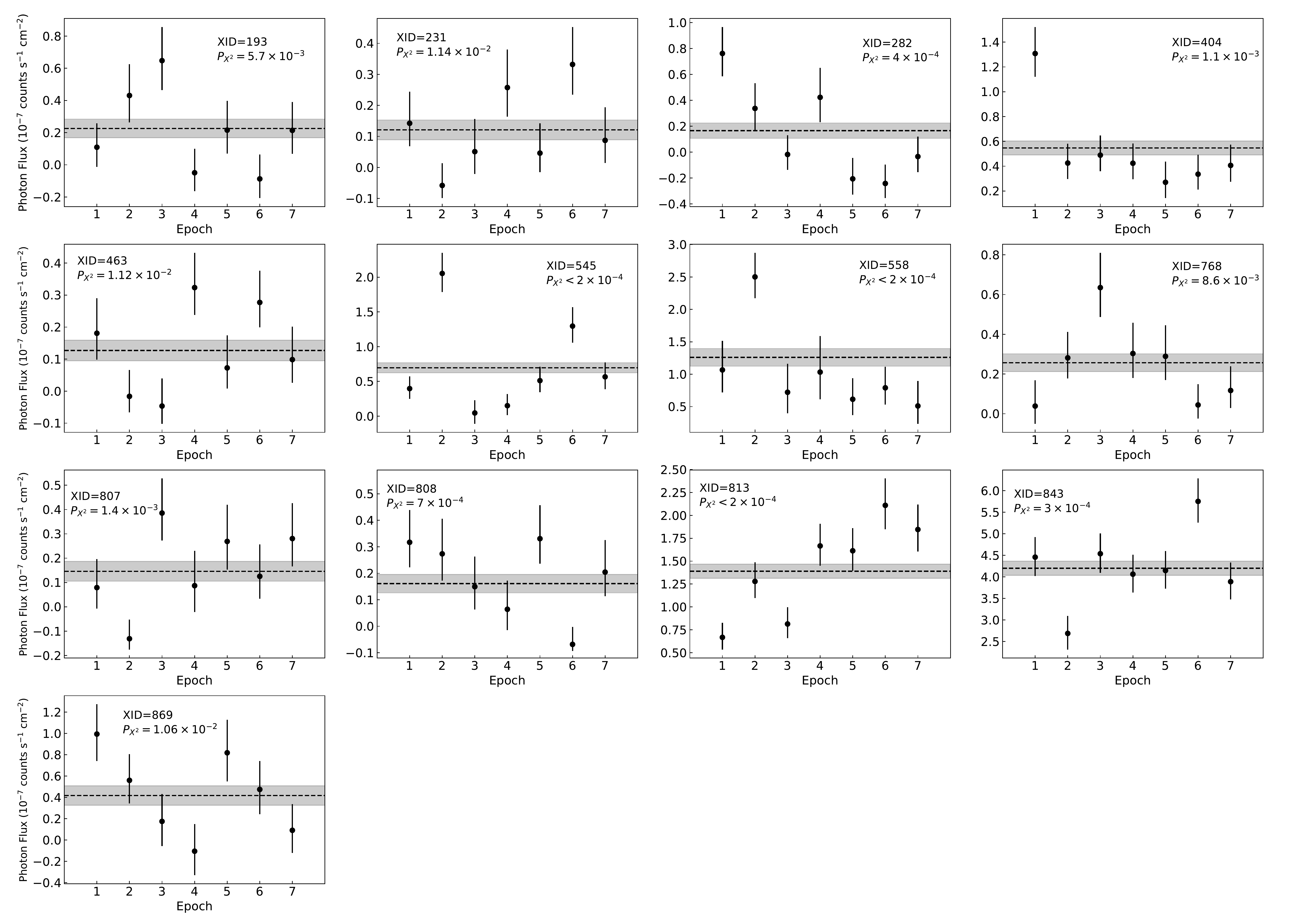}
\caption{Light curves for the 13 variable \mbox{X-ray} sources. In each panel, the black dashed line is the photon flux of the source from the stacked 7~Ms data, and the gray shaded area represents its 1$\sigma$ error. The labels in each panel are the source sequence number (XID) in the 7~Ms main catalog and the $P_{X^2}$ value. Note that the seven epochs are not spaced evenly in time (see~Table~\ref{Table1}).}
\label{fig:VG}
\end{figure*}

The $P_{X^2}$ values of the 110 unclassified CDF-S \mbox{X-ray} sources are calculated using the above method. We adopt $P_{X^2} < 0.015$ as the threshold for selecting variable sources. This conservative selection criterion (compared to $P_{X^2} < 0.05$ in \citealt{2012ApJ...748..124Y}) is chosen to balance the need of recovering a significant number of true variable sources while keeping the number of false-positive sources relatively small. We find 13 variable sources from the 110 objects (see Table~\ref{Table2}), of which $\sim$ 1--2 variable sources are theoretically expected to be false-positive according to the selection criterion ($P_{X^2} < 0.015$). The 13 variable sources account for $\approx12^{+4}_{-2}\%$ of the total number of the unclassified X-ray sources, where the binomial errors on the fraction at a $1\sigma$ confidence level are calculated following \citet{2011PASA...28..128C}.
The light curves of the 13 variable sources are shown in Figure~\ref{fig:VG}.

For completeness, we also list five lower-significance variable sources with $0.015 < P_{X^2} < 0.05$ in Table~\ref{Table3}. Statistically, we expect that $\sim$ 3--4 sources among these five objects are \mbox{false-positive} detections. In the following analyses, we do not treat these five objects as variable sources.

\begin{table*}
\tablenum{2}
\centering
\caption{Properties of the 13 variable CDF-S \mbox{X-ray} sources}
\scriptsize
\begin{tabular}{@{}ccccccccccc@{}}
\hline\hline
XID	&	$z$	&	Net 	&	$\log L_{\mathrm{0.5-7keV, int}}$	&	$X^2$	&	$P_{X^2}$	& $\sigma^2_{\mathrm{NXS, corr}}$&SFR&	$\log M_{\star}$  & $\Gamma_\mathrm{eff}$  \\
	&				&	   counts	&	(erg~s$^{-1})$	& &		&	  &	($M_{\sun}$~yr$^{-1}$)	&	($M_{\sun}$) \\
(1)	&	(2)	&	(3)	&	(4)	&	(5)	&	(6)	&	(7)	&	(8)	&	(9)	& (10) \\
\hline
193	&	0.085	&	57.6	&	39.5	&	18.1	&	0.0057	&	0.57	$\pm$	0.45	&	1.41	&	9.3	&	1.4			\\
231	&	0.333	&	32.3	&	40.5	&	17.6	&	0.0114	&	0.51	$\pm$	0.45	&	0.44	&	8.7	&	1.6			\\
282	&	0.674	&	40.6	&	41.5	&	35.6	&	0.0004	&	4.43	$\pm$	1.81	&	4.68	&	10.2	&	3.0			\\
404	&	0.131	&	138.4	&	40.3	&	22.7	&	0.0011	&	0.34	$\pm$	0.27	&	0.28	&	9.0	&	$1.24_{-0.19}^{+0.21}$			\\
463	&	0.834	&	34.0	&	41.5	&	17.9	&	0.0112	&	0.64	$\pm$	0.38	&	11.12	&	9.4	&	2.6			\\
545	&	0.668	&	175.8	&	42.1	&	61.0	&	$<0.0002$	&	1.28	$\pm$	0.61	&	45.71	&	10.1	&	$1.12_{-0.20}^{+0.21}$		\\
558	&	0.075	&	119.5	&	39.8	&	27.8	&	$<0.0002$	&	0.21	$\pm$	0.13	&	5.49	&	9.9	&	$1.88_{-0.25}^{+0.25}$		\\
768	&	0.147	&	63.3	&	39.9	&	17.5	&	0.0086	&	0.34	$\pm$	0.28	&	0.33	&	8.9	&	$2.01_{-0.50}^{+0.51}$		\\
807	&	0.496	&	36.7	&	40.9	&	26.8	&	0.0014	&	0.23	$\pm$	0.53	&	0.12	&	10.8	&	1.8			\\
808	&	0.738	&	43.5	&	41.4	&	33.1	&	0.0007	&	0.05	$\pm$	0.31	&	12.88	&	9.9	&	2.1		\\
813	&	0.414	&	369.1	&	41.8	&	48.9	&	$<0.0002$	&	0.12	$\pm$	0.04	&	7.41	&	10.2	&	$1.53_{-0.11}^{+0.11}$		\\
843	&	0.103	&	983.5	&	40.8	&	25.7	&	0.0003	&	0.03	$\pm$	0.02	&	0.09	&	11.0	&	$1.98_{-0.10}^{+0.09}$		\\
869	&	0.077	&	101.0	&	39.7	&	15.7	&	0.0106	&	0.34	$\pm$	0.26	&	0.05	&	8.2	&	1.5		\\
\hline
\end{tabular}
\tablecomments{Column(1): Sequence number in the 7~Ms CDF-S main catalog \citep{2017ApJS..228....2L}. Column(2): Spectroscopic redshift. Column(3): Total number of net counts in 0.5--7 keV band. Column(4): Logarithm of the intrinsic rest-frame 0.5--7 keV luminosity from \citet[][]{2017ApJS..228....2L}. Column(5): Variability statistic. Column(6): Probability that the measured variability is due to Possion noise alone. Column(7): Normalized excess variance corrected for sampling bias and its $1\sigma$ uncertainty; Column(8): Star formation rate from \citet[][]{2015ApJ...801...97S}. Column(9): Logarithm of stellar mass from \citet[][]{2015ApJ...801...97S}. Column(10): Effective power-law photon index and its 1$\sigma$ lower and upper uncertainties from \citet[][]{2017ApJS..228....2L}. For source detected only in the soft or hard bands, a best-guess value of the effective power-law photon index was estimated from BEHR, and it has no error estimate.}
\label{Table2}
\end{table*}

\subsection{Comparison with previous results}
\begin{figure}[h]
\centering
\includegraphics[width=3in]{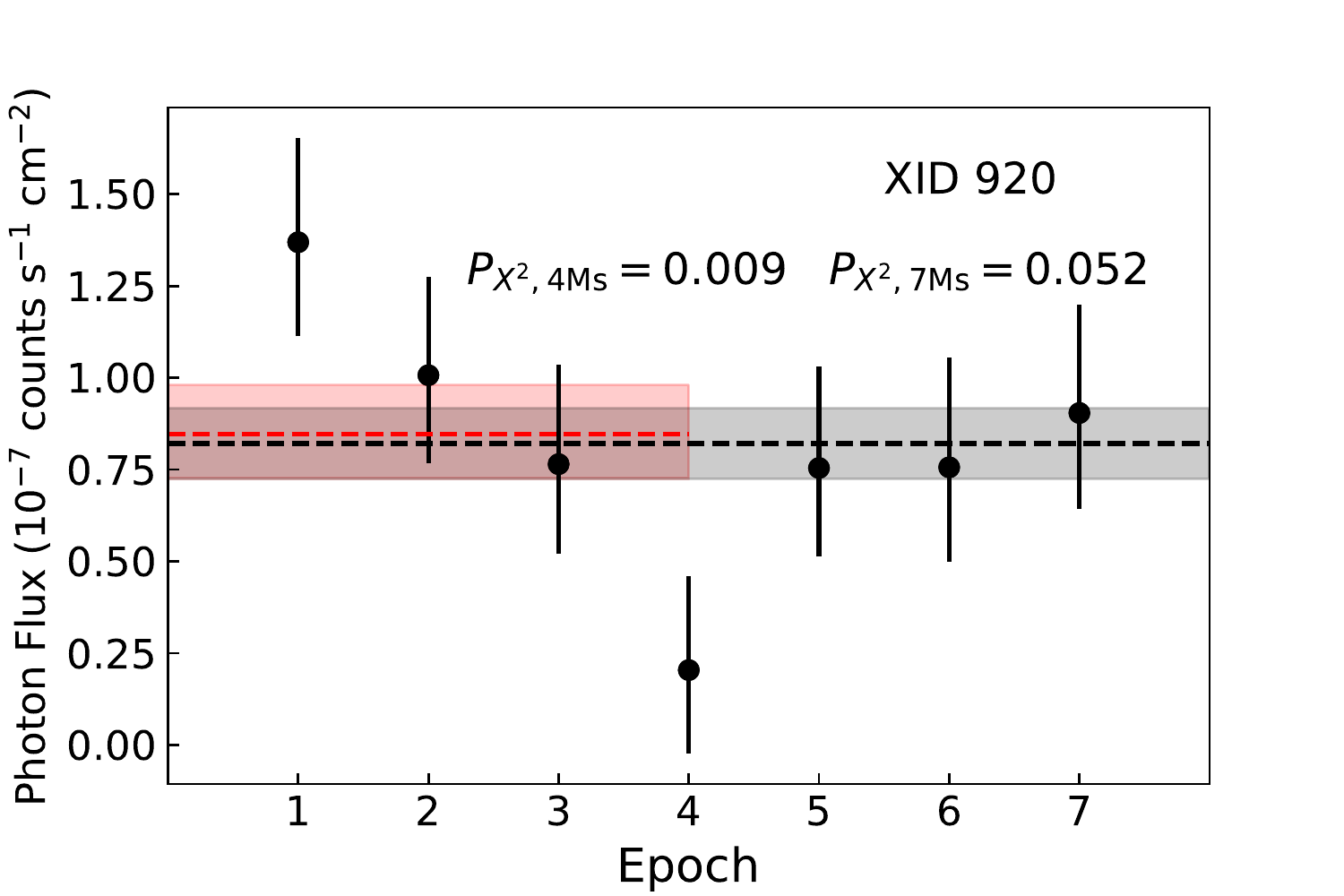}
\caption{Light curve of XID 920. The black dashed line is the photon flux from the stacked 7~Ms (all seven epochs) data, and the gray shaded area represents its 1$\sigma$ error. The red dashed line is the photon flux from the stacked 4~Ms (the first four epochs) data, and the red shaded area represents its 1$\sigma$ error. The labels are XID, the $P_{X^2, \mathrm{4~Ms}}$ value of the 4~Ms data, and the $P_{X^2}$ value of the 7~Ms data. The photon flux of this source exhibits significant fluctuation during the first four epochs. However, in subsequent epochs (\mbox{5--7}), the photon flux of this source remains almost constant; the probability that the photon-flux fluctuation is produced by random noise alone increases and no longer meets the probability threshold requirement for selecting variable sources.}
\label{fig:example}
\end{figure}

\citet{2012ApJ...748..124Y} utilized the same method to search for variable \mbox{X-ray} sources in the 4~Ms CDF-S \citep{2011ApJS..195...10X}. They found 20 variable sources from 92 unclassified CDF-S X-ray sources using the probability threshold of \mbox{$P_{X^2} < 0.05$}. One of these 20 sources (XID~319) is actually a star and it is correctly classified in \citet{2017ApJS..228....2L} based on its spectroscopy. Among the remaining 19 sources\footnote{The 19 sources are all unclassified in \citet{2017ApJS..228....2L}.}, we confirm that eight sources\footnote{Their XIDs are 193, 282, 404, 558, 768, 813, 843, and 869.} do have \hbox{X-ray} variability (\mbox{$P_{X^2} < 0.015$}). 
However, for the other 11 sources, our results show that when considering the 7~Ms \mbox{CDF-S} data, they do not satisfy the \mbox{$P_{X^2} < 0.015$} or even the \mbox{$P_{X^2} < 0.05$} threshold. We examine the light curves of the 11 sources and find the following two cases:

1) If only the 4~Ms CDF-S data (i.e., the first four epochs of data) are considered, there are eight variable sources\footnote{Their XIDs are 206, 298, 323, 349, 371, 565, 750, and 920.} whose $P_{X^2}$ values meet the probability threshold of $P_{X^2} < 0.05$ used in \citet{2012ApJ...748..124Y}.
However, since the photon fluxes of these sources have hardly changed in subsequent epochs (5--7), when the 7~Ms \mbox{CDF-S} data (i.e., all seven epochs) are considered, the $P_{X^2}$ values of these sources no longer meet the probability threshold requirement for selecting variable sources ($P_{X^2} < 0.05$). For example, Figure~\ref{fig:example} displays the light curve of one of the sources, \hbox{XID 920}. The photon flux of this source exhibits significant fluctuation during the first four epochs (i.e., 4~Ms data; $P_{X^2, \mathrm{4Ms}} = 0.009$). However, in subsequent epochs (5--7), the photon flux of this source remains almost constant; the probability that its photon-flux fluctuation is produced by random fluctuation alone increases to $P_{X^2, \mathrm{7Ms}} = 0.052$, and it no longer satisfies the probability threshold requirement for selecting variable sources. There is a possibility that such sources are actually \hbox{X-ray} variable, but in the latest three epochs of observations, they do not display significant variability due to relatively short time separations between the epochs, producing a larger $P_{X^2}$ value than that from the 4~Ms data. For each of these eight sources, we perform simulations to estimate the probability that its variability cannot be detected with the 7~Ms data (i.e., $P_{X^2, \mathrm{7Ms}} > 0.05$), assuming that it is a variable source and its variability had been detected when considering the 4~Ms data (i.e., $P_{X^2, \mathrm{4Ms}} < 0.05$). The probabilities for the eight sources range between $\approx \textrm{28--39}\%$ (see Section 5.1 below for details). This result implies that $\sim\textrm{2--3}$ sources among the eight may possess \hbox{X-ray} variability, although their variability cannot be detected when considering the 7~Ms \mbox{CDF-S} data. For completeness, we list the eight sources in Table~4 as possible candidates for being \mbox{X-ray} variable. Statistically, $\sim\textrm{5--6}$ of these sources are expected to be \mbox{false-positive} detections. We do not treat these objects as variable sources in the following analyses.

2) For the other three sources (XIDs 149, 690, 916), their $P_{X^2}$ values do not meet the probability threshold of $P_{X^2} < 0.05$ considering either the first four epochs of data or all seven epochs of data. The different results from \citet{2012ApJ...748..124Y} are mainly due to the differences in the extracted \mbox{X-ray} photometry. 
Compared to the 4~Ms survey, we detected fainter \mbox{X-ray} sources in the 7~Ms survey which influences the estimated background levels, and for existing sources, we detected larger numbers of counts. These allow us to obtain more accurate source positions and photometric properties with the 7~Ms data.

In summary, among the 20 variable sources found in \citet{2012ApJ...748..124Y}, XID~319 is actually a star, and 16 of the remaining 19 sources, using our extraction process, meet the selection criterion of variable sources ($P_{X^2} < 0.05$) used in \citet{2012ApJ...748..124Y} if considering the 4~Ms \mbox{CDF-S} data alone. However, only eight of the 16 sources satisfy the selection criterion of variable sources ($P_{X^2} < 0.05$) when using the 7~Ms \mbox{CDF-S} data.

Our analysis identifies five new variable sources (XIDs 231, 463, 545, 807, 808) relative to \citet{2012ApJ...748..124Y}; the details of these sources are as follows:

XID 231: The source is not included in the sample of \citet{2012ApJ...748..124Y} because it does not have sufficient photons (less than 20 net counts in the first four epochs; see \citealt{2012ApJ...748..124Y}).

XID 545: The luminosity of the source is close to the boundary of AGN identification (\mbox{$L_{\mathrm{X}} >3 \times 10^{42}~\mathrm{erg~s}^{-1}$}), which causes this source to be classified into different types in the 4~Ms \mbox{CDF-S} and 7~Ms \mbox{CDF-S} (the source is classified as an AGN in the 4~Ms \mbox{CDF-S}; see \citealt{2011ApJS..195...10X}). Here, we find that this source exhibits significant \hbox{X-ray} variability, which is consistent with its identification in the 4~Ms \mbox{CDF-S}.

XIDs 463, 807: These two objects meet the selection criterion for variable sources ($P_{X^2} < 0.015$) considering either the first four epochs of data or all seven epochs of data. 
The different results from \citet{2012ApJ...748..124Y} are mainly due to the differences in the extracted \mbox{X-ray} photometry, and we expect that the photometric properties from the 7~Ms data are more accurate as explained in point 2) above.

XID 808: The source does not meet the selection criterion of variable sources ($P_{X^2} < 0.05$) when considering the first four epochs of data. In the subsequent three epochs (5--7), however, the photon fluxes of this source show significant change. When considering all seven epochs of data, the source meets the selection criterion of variable sources ($P_{X^2} < 0.015$).

We also compare our selection to AGN candidates selected by optical variability in the \mbox{CDF-S}.
\citet{2008A&A...488...73T} utilized optical variability (timescale of $\sim 3$~years) to search for AGN candidates in the AXAF field ($\sim0.25$~deg$^2$) that overlaps the \mbox{CDF-S}. They reported 132 optically variable AGN candidates, 39 of which fall into the field of view of the CDF-S. A total of 23 of the 39 sources have \mbox{X-ray} counterparts in the 7~Ms~CDF-S main catalog using a matching radius of 0.5$''$. Among these 23 objects, 22 have been classified as AGNs in \citet{2017ApJS..228....2L}, and only one source (XID 558) is unclassified. Our variability calculation suggests that XID 558 has significant \mbox{X-ray} variability, which is consistent with the result of \citet{2008A&A...488...73T} that XID 558 could be an AGN. 

\citet{2010ApJ...723..737V} utilized optical variability (timescale of $\sim \textrm{3--6}$~months) to search for AGN candidates in the GOODS fields ($\sim0.1$~deg$^2$), and the GOODS-South field overlaps the \mbox{CDF-S}. They reported 139 optically variable AGN candidates, of which 40 high-significance (a significance level of $99.99\%$) candidates fall into the field of view of the CDF-S. A total of 19 of the 40 objects have \mbox{X-ray} counterparts in the 7~Ms~CDF-S main catalog using a matching radius of 0.5$''$, of which 12 sources have been classified as AGNs in \citet{2017ApJS..228....2L} and seven sources are unclassified. Four (XIDs~446, 498, 553, and 558) of these seven unclassified sources are included in our sample (i.e., 110 unclassified \mbox{X-ray} sources). As mentioned above, XID 558 has significant \mbox{X-ray} variability. Our variability calculation suggests that XID~553 has \mbox{lower-significance} \mbox{X-ray} variability (see Table~\ref{Table3}). XID~446 and 498 are not found to have \mbox{X-ray} variability, which is probably because:
a) they have small numbers of net counts (XIDs~446 and 498 only have 45 and 32 total observed net counts, respectively), making their variability difficult to detect; or b) they do not display \mbox{X-ray} variability during the \mbox{CDF-S} survey.
Among the remaining 21 ($40-19$) candidates without \mbox{X-ray} counterparts, 11 have off-axis angles of $<6'$ in the 7~Ms \mbox{CDF-S} and they are not adjacent to any \mbox{X-ray} sources. We perform a stacking analysis for these 11 sources using the 7 Ms \mbox{CDF-S} data, and this sample does not show any stacked X-ray signal with a $3\sigma$ flux upper limit of $3.9\times10^{-18}$ erg~cm$^{-2}$~s$^{-1}$ in the 0.5--2~keV band (corresponding to a luminosity upper limit of $4.4\times10^{39}$~erg~s$^{-1}$ at a mean redshift of 0.52). There are two possible interpretations regarding the nature of these objects: a) they are not AGNs in general but are other kind of objects such as nuclear supernovae (e.g., \citealt{2005NewAR..49..430B}); b) they are AGNs with small-scale X-ray absorbers which obscure X-ray emission but do not affect the observed optical emission (e.g., similar to the obscuring material in broad absorption line quasars).

\citet{2015A&A...579A.115F} utilized optical variability (timescale of $\sim 1$~year) to search for AGN candidates in the SUDARE-VOICE field ($\sim2$~deg$^2$) that covers the \mbox{CDF-S}. They reported 175 optically variable AGN candidates, only eight of which fall into the field of view of the CDF-S. Four of the eight sources have \mbox{X-ray} counterparts in the 7~Ms~CDF-S main catalog with a matching radius of 0.5$''$, and all four have been classified as AGNs in \citet{2017ApJS..228....2L}. 

\section{LLAGN CANDIDATES}
There are three possible scenarios that may be responsible for the \mbox{X-ray} variability of the 13 variable sources, namely \mbox{X-ray} binary (XRB) populations, ultraluminous \mbox{X-ray} sources (ULXs), and accreting SMBHs. In this section, we investigate which scenario is the most probable case. For XID 869, its \mbox{X-ray} variability may be due to a ULX, while for the vast majority of the remaining 12 variable sources, their \mbox{X-ray} variability is most likely explained by the scenario of accreting SMBHs (i.e., AGNs).

\subsection{Measuring variability amplitude}
In order to analyze quantitatively the characteristics of variability, we adopt the normalized excess variance defined in \citet{2003MNRAS.345.1271V}
\begin{equation}
\label{eq7}
\begin{split}
\sigma^2_\mathrm{NXS} =  \frac{1}{(N-1)\langle PF \rangle^2}\sum_{i=1}^N(PF_{i} - \langle PF \rangle)^2\\
-\frac{1}{N\langle PF \rangle^2}\sum_{i=1}^N(\delta PF_i)^2.
\end{split}
\end{equation}

The statistical uncertainty of $\sigma^2_\mathrm{NXS}$ is computed as $S_D/(\langle PF \rangle^2\sqrt{N})$ \citep{2003MNRAS.345.1271V}, where
\begin{equation}
\label{eq8}
S_D^2 = \frac{1}{N-1}\sum_{i=1}^N[(PF_{i} - \langle PF \rangle)^2 - (\delta PF_{i})^2-\sigma^2_\mathrm{NXS}\langle PF \rangle^2]^2.
\end{equation}

The normalized excess variance measures how strongly the photon flux fluctuation of a source exceeds the expected measurement error. When photon flux fluctuation is entirely consistent with noise rather than due to intrinsic source variability, the normalized excess variance is consistent with zero (i.e., $\sigma^2_\mathrm{NXS} = 0$). Due to statistical fluctuations, the normalized excess variance may be negative. The normalized excess variance is calculated for an observed-frame energy band. For sources with different redshifts, their $\sigma^2_\mathrm{NXS}$ values actually reflect the variability amplitude of photon fluxes in different rest-frame energy bands. Nevertheless, there is evidence suggesting that the bandpass effects are small both for \mbox{short-term} or \mbox{long-term} variability of AGNs \citep[e.g.,][]{2012A&A...542A..83P, 2016ApJ...831..145Y, 2017arXiv171004358Z}. Thus, we do not expect that the bandpass effects will materially affect our following analyses.

In practical measurements, the $\sigma^2_\mathrm{NXS}$ value will be affected by the sparse observing pattern of the CDF-S survey. Because the measured mean of the photon fluxes of a source ($\mu$) reflects the mean of its sampled data points rather than the true mean of its photon fluxes (i.e., $\langle PF \rangle$), the measured variance of its photon flux fluctuation ($\sigma^2 = \frac{1}{N-1}\sum_{i=1}^N(PF_{i} - \mu)^2$) is not the true variance (i.e., $\sigma^2_{\mathrm{true}} = \frac{1}{N-1}\sum_{i=1}^N(PF_{i} - \langle PF \rangle)^2$), and its $\sigma^2$ value (as well as $\sigma^2_\mathrm{NXS}$ value) will tend to be underestimated (e.g., \citealt{2012ApJ...748..124Y, 2013ApJ...771....9A}). Therefore, for each source, we apply a Monte Carlo simulation method to correct the measured mean ($\mu$) of photon fluxes and the measured variance ($\sigma^2$) of photon flux fluctuation, respectively, and then calculate the bias-corrected $\sigma^2_\mathrm{NXS}$ value (i.e., $\sigma^2_\mathrm{NXS, corr}$; e.g., \citealt{2012ApJ...748..124Y}). The specific procedure is described as follows:

we assume that all sources have intrinsic \mbox{X-ray} variability. For non-variable sources, the measured variance of photon flux fluctuation is regarded as an upper limit to the variability that could be present. We assume that the intrinsic \mbox{X-ray} variability of all sources can be simply described by a power-law power spectral density (PSD) function (i.e., $P(f)\propto f^{-\beta}$).\footnote{We compare the $\sigma^2_\mathrm{NXS, corr}$ values calculated using this simple power-law PSD function to those calculated using a more complicated PSD function described in Section 5.4 below. We find that the differences between the $\sigma^2_\mathrm{NXS, corr}$ values calculated using these two different PSD functions are small, far less than the typical uncertainty of $\sigma^2_\mathrm{NXS, corr}$. Therefore, for simplicity, this simple \mbox{power-law} PSD function is used in the calculations of $\sigma^2_\mathrm{NXS, corr}$.} This PSD has been usually adopted in previous studies \citep[e.g.,][]{2012ApJ...748..124Y,2013ApJ...771....9A}. Previous investigations of the long-term variability of nearby Seyfert galaxies found that $\beta\sim1$ \citep[e.g.,][]{2002MNRAS.332..231U, 2003MNRAS.341..496V, 2003MNRAS.339.1237V}. However, recent studies reported that $\beta$ may not be $1$ and suggested $\beta = \textrm{1.2--1.3}$ \citep[e.g.,][]{2017arXiv171004358Z}. We adopt $\beta=1$ and $\beta=1.3$ to perform subsequent calculations, respectively. Following the procedure in \mbox{Section 3} of \citet{2013ApJ...771....9A}, we take measured mean ($\mu$) and variance ($\sigma^2$) as ``true'' values to generate 5000 stochastic light curves based on the given power-law PSD, where each simulated light curve is generated five times longer than the sampled region in order to reproduce the effect of  ``red-noise leak''. Each simulated light curve is resampled based on the actual observing pattern, and the sparsely sampled light curve is modified with Poisson noise to account for measurement errors. Finally, the mean ($\mu_\mathrm{sim}$ ) and variance ($\sigma^2_\mathrm{sim}$) of each simulated light curve is calculated, and the ratio of the ``true'' input value (i.e., the measured values used as input in the simulations) and the median of output values of all simulated light curves (i.e., the biased values produced by sparse sampling) are computed as

\begin{equation}
\label{eq9}
f_\mathrm{mean} = \mu/\mathrm{median}(\mu_\mathrm{sim}),
\end{equation}
\begin{equation}
\label{eq10}
f_\mathrm{variance} = \sigma/\mathrm{median}(\sigma_\mathrm{sim}),
\end{equation}
where $f_\mathrm{mean}$ is the rescaling factor of the mean and $f_\mathrm{variance}$ is the rescaling factor of the variance. For our full sample (505 sources), the values of $f_\mathrm{mean}$ range between $\approx\textrm{0.6--1.2}$, with a median value of $\approx0.9$. The values of $f_\mathrm{variance}$ vary between $\approx\textrm{0.5--2.7}$, with a median value of $\approx1.2$. For $69\%$ of the sources, their $f_\mathrm{variance}$ values are larger than one, indicating that their variances are underestimated due to the sampling bias. Using the rescaling factors, the biased measured mean ($\mu$) and variance ($\sigma^2$) can be corrected (i.e., \mbox{$\langle PF \rangle = f_\mathrm{mean} \times \mu$}; \mbox{$\sigma_{\mathrm{true}} = f_\mathrm{variance} \times \sigma$}), and then the bias-corrected $\sigma^2_\mathrm{NXS}$ value ($\sigma^2_\mathrm{NXS, corr}$) and its statistical uncertainty ($\mathrm{err}(\sigma^2_\mathrm{NXS, corr}$)) are calculated using Eqs. 7--8.

We perform the above procedure to calculate the $\sigma^2_\mathrm{NXS, corr}$ value of each source in our full sample (505 sources). The influence of different $\beta$ values adopted (i.e., $\beta = 1$ and $\beta = 1.3$) on the calculated results of $\sigma^2_\mathrm{NXS, corr}$ is small, with an average relative deviation of $\sim3.2\%$. This difference is much smaller than the statistical uncertainty of $\sigma^2_\mathrm{NXS, corr}$ itself. Thus, for simplicity, we only use $\sigma^2_\mathrm{NXS, corr}$ values calculated with $\beta = 1$ in the following analyses. The $\sigma^2_\mathrm{NXS, corr}$ values and their $1\sigma$ uncertainties for the 13 variable sources are listed in Table 2, where
the $\sigma^2_\mathrm{NXS, corr}$ values of two variable sources (XID 807 and XID 808) are completely dominated by statistical uncertainty (i.e., $\sigma^2_\mathrm{NXS, corr}\leq\mathrm{err}(\sigma^2_\mathrm{NXS, corr}$)). Even though we correct the bias produced by sparse sampling, an individual $\sigma^2_\mathrm{NXS, corr}$ value may still be highly uncertain, as shown in \citet[][]{2013ApJ...771....9A}. Thus the $\sigma^2_\mathrm{NXS, corr}$ values are best considered in the ensemble of a specific group rather than on an individual object basis.

\begin{table*}
\tablenum{3}
\centering
\caption{Properties of the five lower-significance variable sources with $0.015<P_{X^2}<0.05$}
\scriptsize
\begin{tabular}{@{}ccccccccccc@{}}
\hline\hline
XID	&	$z$	&	Net 	&	$\log L_{\mathrm{0.5-7keV, int}}$	&	$X^2$	&	$P_{X^2}$	& $\sigma^2_{\mathrm{NXS, corr}}$&SFR&	$\log M_{\star}$  & $\Gamma_\mathrm{eff}$  \\
	&				&	   counts	&	(erg~s$^{-1})$	& &		&	  &	($M_{\sun}$~yr$^{-1}$)	&	($M_{\sun}$)  \\
(1)	&	(2)	&	(3)	&	(4)	&	(5)	&	(6)	&	(7)	&	(8)	&	(9)	& (10) \\
\hline
80	&	0.575	&	224.2	&	42.0	&	12.3	&	0.0373	&	0.01	$\pm$	0.07	&	...	&	...	&	1.5		\\
219	&	0.576	&	40.2	&	41.1	&	12.0	&	0.0444	&	0.26	$\pm$	0.33	&	34.88	&	10.2	&	2.0		\\
553	&	0.076	&	289.0	&	40.0	&	12.8	&	0.0336	&	0.01	$\pm$	0.02	&	6.03	&	10.0	&	$2.62^{+0.26}_{-0.28}$		\\
663	&	0.542	&	33.0	&	41.0	&	16.1	&	0.0215	&	-0.17	$\pm$	0.24	&	10.23	&	9.7	&	1.4			\\
825	&	0.438	&	56.1	&	40.9	&	12.9	&	0.0337	&	0.11	$\pm$	0.14	&	14.45	&	10.1	&	1.8		\\
\hline
\end{tabular}
\tablecomments{The same format as Table~\ref{Table2} but for the five lower-significance variable sources with $0.015<P_{X^2}<0.05$. The measurements of the star formation rate and stellar mass are not available for XID 80.
Statistically, $\sim\textrm{3--4}$ sources among the five objects are expected to be \mbox{false-positive} detections. See Section~3.2 for details.}
\label{Table3}
\end{table*}

\begin{table*}
\tablenum{4}
\centering
\caption{Properties of the eight \mbox{X-ray} variable candidates selected based on $P_{X^2, \mathrm{4Ms}} < 0.05$}
\scriptsize
\begin{tabular}{@{}ccccccccccccc@{}}
\hline\hline
XID	&	$z$	&	Net 	&	$\log L_{\mathrm{0.5-7keV, int}}$	&	$X^2_{\mathrm{4Ms}}$&  $P_{X^2, \mathrm{4Ms}}$ &  $X^2_{\mathrm{7Ms}}$	  &	$P_{X^2, \mathrm{7Ms}}$  & $\sigma^2_{\mathrm{NXS, corr}}$ & SFR&	$\log M_{\star}$  & $\Gamma_\mathrm{eff}$  \\
	&				& 	   counts	&	(erg~s$^{-1})$	& &		&	  &&&	($M_{\sun}$~yr$^{-1}$)	&	($M_{\sun}$)  \\
(1)	&	(2)	&	(3)	&	(4)	&	(5)	&	(6)	&	(7)	&	(8)	&	(9)	& (10) & (11) & (12)  \\
\hline
206	&	0.418	&	127.1	&	41.6	&	6.2	&	0.0312	&	9.5	&	0.1066	&	-0.01	$\pm$	0.12	&	0.81	&	11.0	&	$1.07^{+0.33}_{-0.31}$	\\
298	&	0.524	&	102.0	&	41.6	&	4.2	&	0.0411	&	5.4	&	0.4065	&	-0.11	$\pm$	0.07	&	0.05	&	9.8	&	$1.28^{+0.34}_{-0.33}$	\\
323	&	0.734	&	79.5	&	41.7	&	6.4	&	0.0111	&	9.8	&	0.0854	&	0.01	$\pm$	0.05	&	0.25	&	11.7	&	$2.54^{+0.39}_{-0.43}$	\\
349	&	0.964	&	207.4	&	42.4	&	9.2	&	0.0224	&	7.3	&	0.2331	&	-0.03	$\pm$	0.04	&	0.47	&	11.4	&	$2.30^{+0.37}_{-0.37}$	\\
371	&	0.679	&	30.1	&	41.1	&	5.0	&	0.0442	&	7.1	&	0.2098	&	0.01	$\pm$	0.48	&	47.79	&	10.2	&	2.1	\\
565	&	0.648	&	62.1	&	41.4	&	7.9	&	0.0131	&	7.8	&	0.1746	&	-0.08	$\pm$	0.10	&	10.08	&	9.5	&	2.3	\\
750	&	0.522	&	25.0	&	40.8	&	6.4	&	0.0483	&	8.6	&	0.1350	&		...		&	...	&	...	&	1.9	\\
920	&	0.104	&	202.4	&	40.2	&	10.5	&	0.0090	&	11.5	&	0.0522	&	0.02	$\pm$	0.09	&	...	&	...	&	$1.61^{+0.28}_{-0.27}$	\\
\hline
\end{tabular}
\tablecomments{Column(5) and Column(7) are the variability statistics from the first four and all seven epoch data, respectively. Column(6) and Column(8) are probabilities that the measured variability is due to Possion noise alone considering the first four epoch data and all seven epoch data, respectively. The other columns have the same format as Table~\ref{Table2}. XID~750 is included in the sample of \citet{2012ApJ...748..124Y}, but it is not in our sample because it does not meet our sample selection criterion (less than 30 net counts in the full 7 Ms exposure). The measurements of the star formation rate and stellar mass are not available for XIDs 750 and 920. Statistically, $\sim\textrm{5--6}$ sources among the eight objects are expected to be \mbox{false-positive} detections. See Section~3.3 for details.}
\label{Table4}
\end{table*}

\subsection{The scenario of XRB population}
Discrete XRBs are the main contributors of \mbox{galaxy-wide} \mbox{X-ray} luminosity for normal galaxies. To diagnose whether XRB populations can be responsible for the \mbox{X-ray} variability of these variable sources, below we compare the measured intrinsic \mbox{X-ray} luminosity ($L_{\mathrm{2-10~keV, int}}$) and variability amplitude of each of the 13 variable sources to those expected from XRB populations, respectively.

The potential \mbox{galaxy-wide} \mbox{X-ray} luminosity from XRB populations ($L_{\mathrm{2-10~keV, XRB}}$) is composed of luminosities of low-mass X-ray binaries (LMXBs) and high-mass X-ray binaries (HMXBs). Previous studies have demonstrated that the total X-ray luminosities from the LMXB population and HMXB population are proportional to the stellar mass ($M_\star$) and star formation rate (SFR) of the host galaxy, respectively \cite[e.g.,][]{2004MNRAS.349..146G, 2004MNRAS.347L..57G, 2010ApJ...724..559L, 2016ApJ...825....7L}. \citet{2016ApJ...825....7L} provided the following redshift-dependent empirical relation:
\begin{equation}
\label{eq11}
\begin{split}
L_{\mathrm{2-10~keV, XRB}} = L_\mathrm{LMXB} + L_\mathrm{HMXB}\\
=\alpha(1+z)^{\gamma}M_{\star}+\beta(1+z)^{\delta}\mathrm{SFR},
\end{split}
\end{equation}
where $\log \alpha = 29.30$, $\log \beta = 39.40$, $\gamma=2.19$ and $\delta=1.02$, which are the best-fit values from the 6~Ms~CDF-S data.

We collect the values of $M_{\star}$ and SFR of each source in our full sample (505 sources) by cross-matching the optical counterpart position of each source to the CANDELS catalog \citep{2015ApJ...801...97S} with a matching radius of 0.5$''$. The CANDELS catalog provides multiple estimated values of $M_{\star}$ and SFR that were calculated by different teams through spectral energy distribution (SED) fittings. Following \citet{2017ApJ...842...72Y}, we adopt the median values of $M_{\star}$ and SFR from five teams (labeled as $2a_{\tau}$, $6a_{\tau}$, $11a_{\tau}$, $13a_{\tau}$, and $14a$ in \citealt{2015ApJ...801...97S}). The five teams adopted the same stellar templates from \citet{2003MNRAS.344.1000B} and the Chabrier IMF when performing the SED fittings. Excluding 41 sources which have different redshifts in the CANDELS catalog and the 7~Ms \mbox{CDF-S} catalog ($|\Delta z|/(1+z_{\mathrm{CANDELS}}) > 0.15$), we obtain the $M_{\star}$ and SFR values for 83 non-variable unclassified X-ray sources, 269 AGNs, and the 13 variable sources. The values of $M_{\star}$ and SFR for the 13 variable sources are listed in Table~2. 
Since the empirical relation of \citet[][]{2016ApJ...825....7L} was obtained under the assumption of the \citet{2001MNRAS.322..231K} IMF, in the calculations of $L_{\mathrm{2-10~keV, XRB}}$ using \mbox{Eq. 11}, we correct the values of $M_{\star}$ and SFR from the Chabrier IMF to the Kroupa IMF following the prescription of \citet{2014ARA&A..52..415M}.
The measured 0.5--7~keV intrinsic \mbox{X-ray} luminosity ($L_{\mathrm{0.5-7~keV, int}}$) and effective photon index ($\Gamma_{\mathrm{eff}}$) of each source are provided in \citet[][]{2017ApJS..228....2L}. Based on the effective photon index of each source, the 2--10~keV intrinsic \mbox{X-ray} luminosity ($L_{\mathrm{2-10~keV, int}}$) is converted from $L_{\mathrm{0.5-7~keV, int}}$.

\begin{figure}[]
\centering
\includegraphics[width=3.5in]{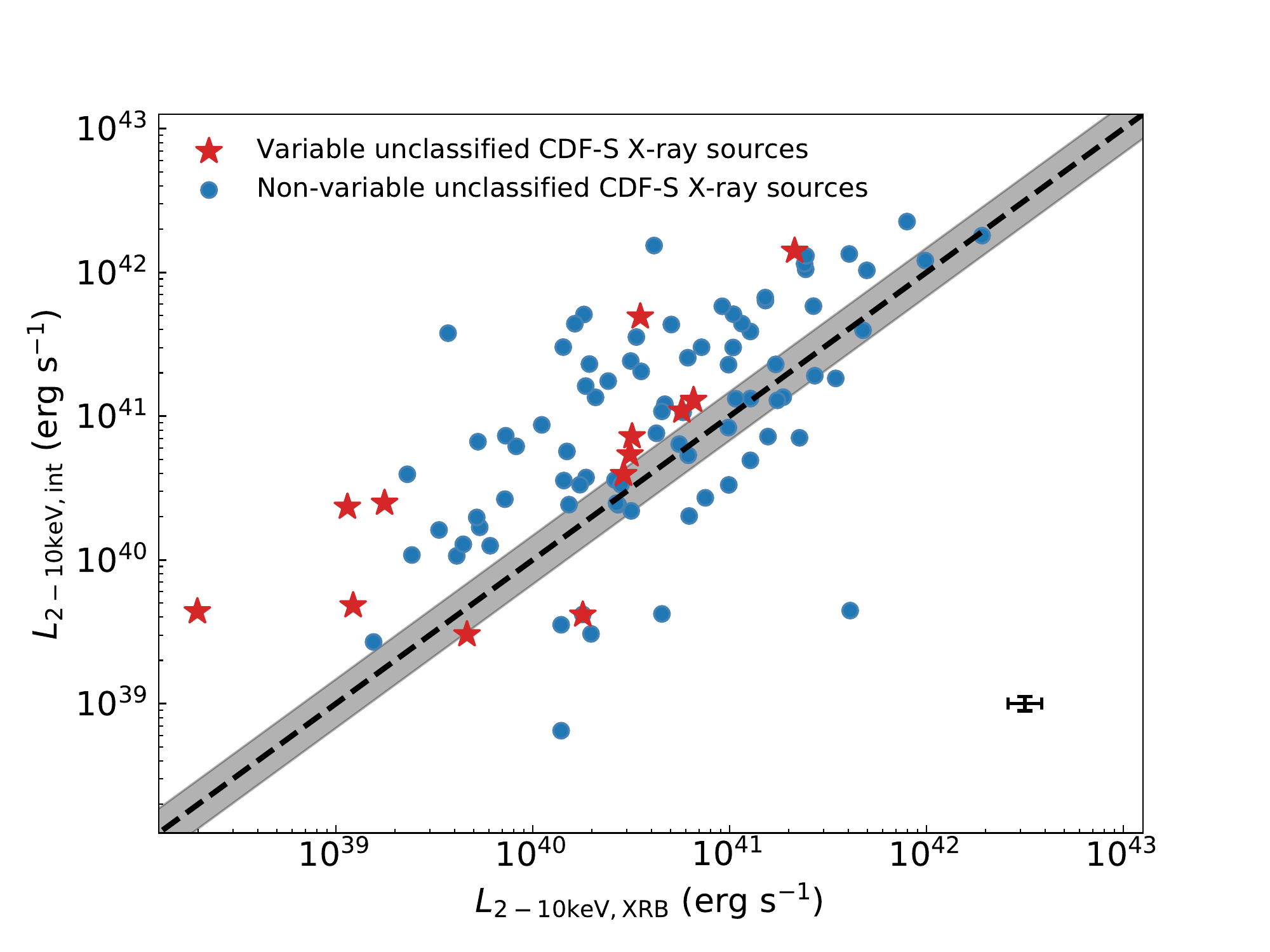}
\caption{Observed 2--10 keV~X-ray luminosity ($L_{\mathrm{2-10~keV, int}}$) vs. that expected from XRB populations ($L_{\mathrm{2-10~keV, XRB}}$) for the 13 variable CDF-S \mbox{X-ray} sources (red filled stars). The data points of the 83 \mbox{non-variable} unclassified \mbox{X-ray} sources with available SFR and $M_{\star}$ measurements (blue filled circles) are plotted for reference. The black dashed line is $L_{\mathrm{2-10~keV, int}} = L_{\mathrm{2-10~keV, XRB}}$, and the gray shaded area shows the expected $1\sigma$ dispersion of the derived $L_{\mathrm{2-10~keV, XRB}}$ (0.17 dex; \citealt{2016ApJ...825....7L}). The black cross in the lower right corner represents the average error bar of the data points.}
\label{fig:L_L}
\end{figure}

Figure~\ref{fig:L_L} presents the measured 2--10~keV intrinsic luminosities versus those expected from XRB populations for the 13 variable sources (red filled stars). The black dashed line indicates the unity relation, while the gray shaded area represents the expected $1\sigma$ dispersion of the derived $L_{\mathrm{2-10~keV, XRB}}$ values (0.17 dex; \citealt{2016ApJ...825....7L}). Except for two variable objects (XIDs 193, 558), the measured luminosities of the remaining variable sources are higher than those expected from XRB populations.
For reference, the data points of the 83 \mbox{non-variable} unclassified \mbox{X-ray} sources with available SFR and $M_{\star}$ measurements are also shown in Figure~\ref{fig:L_L} (blue filled circles), which also generally lie above the unity line. \mbox{Eq. 11} was obtained from stacked samples of normal galaxies in the \mbox{CDF-S}, where most of the galaxies are not detected in the X-ray band \citep[][]{2016ApJ...825....7L}. We also expect strong intrinsic scatter in this empirical relation. Therefore, it is perhaps not surprising to observe that X-ray detected galaxies in the \mbox{CDF-S}, whether variable or \mbox{non-variable}, tend to reside in the bright end of the scatter and lie above the unity relation in Figure~\ref{fig:L_L}. Thus this comparison \mbox{cannot} rule out the possibility that the \mbox{X-ray} emission of the 13 variable sources are produced by XRB populations.

\begin{figure*}[]
\centering
\includegraphics[width=7in]{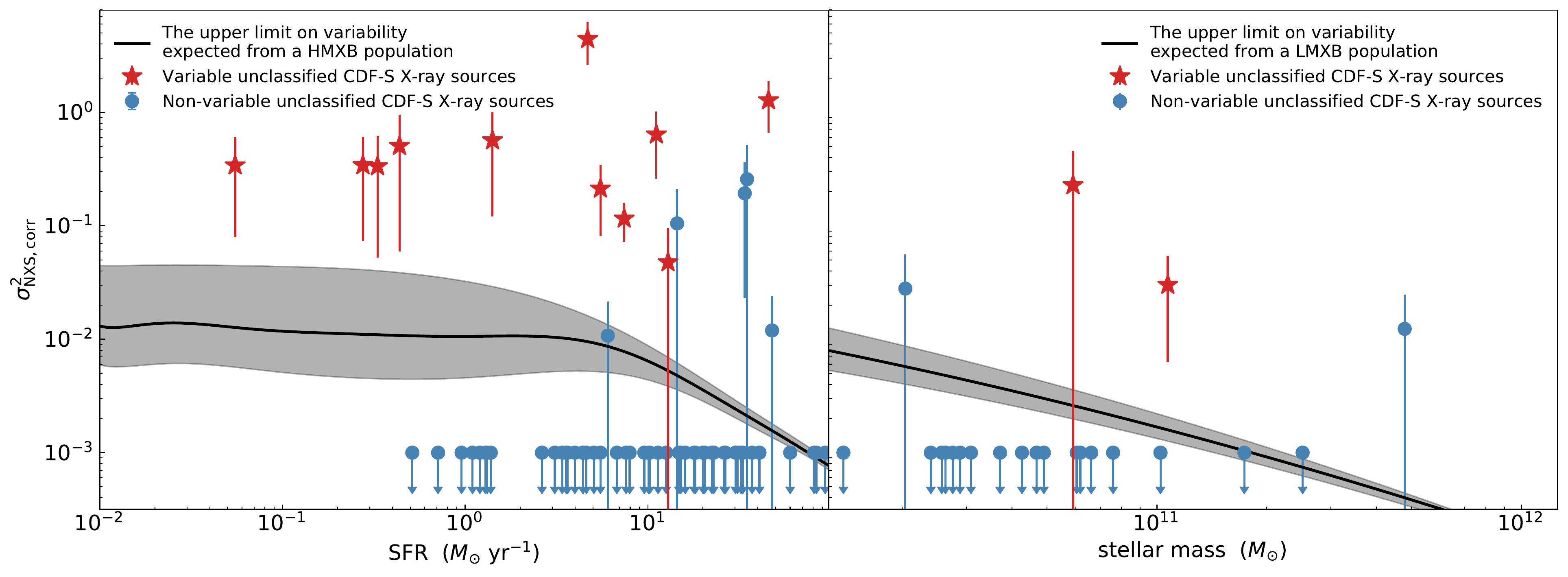}
\caption{Left panel: $\sigma^{2}_{\mathrm{NXS, corr}}$ vs. SFR for the 11 variable CDF-S \mbox{X-ray} sources (red filled stars) whose \mbox{X-ray} luminosities are dominated by a HMXB population. Right panel: $\sigma^{2}_{\mathrm{NXS, corr}}$ vs. $M_{\star}$ for the two variable CDF-S \mbox{X-ray} sources (XIDs 807, 843; red filled stars) whose \mbox{X-ray} luminosities are dominated by a LMXB population. The black solid curve of left (right) panel is the upper limit on $\sigma_{\mathrm{rms,tot}}^{2}$ as a function of SFR ($M_{\star}$) in the case of a HMXB (LMXB) dominant population when adopting $\sigma_{\mathrm{rms,0}} = 30\%$. The shaded areas represent the $1\sigma$ dispersion derived from the simulations of \citet{2004MNRAS.351.1365G}. The data points of the 83 \mbox{non-variable} unclassified \mbox{X-ray} sources with available SFR and $M_{\star}$ measurements (blue filled circles) are also included for reference, where 76 sources with negative $\sigma^2_\mathrm{NXS, corr}$ values are placed at the $10^{-3}$ level and marked by the arrows. The variability amplitudes of the 13 variable sources are significantly higher than the upper limit of variability amplitude expected from XRB populations, which suggest that the variability of these variable sources is not likely caused by XRB populations.}
\label{fig:sigma2_SFR_M}
\end{figure*}

\begin{figure*}[]
\centering
\includegraphics[width=6in]{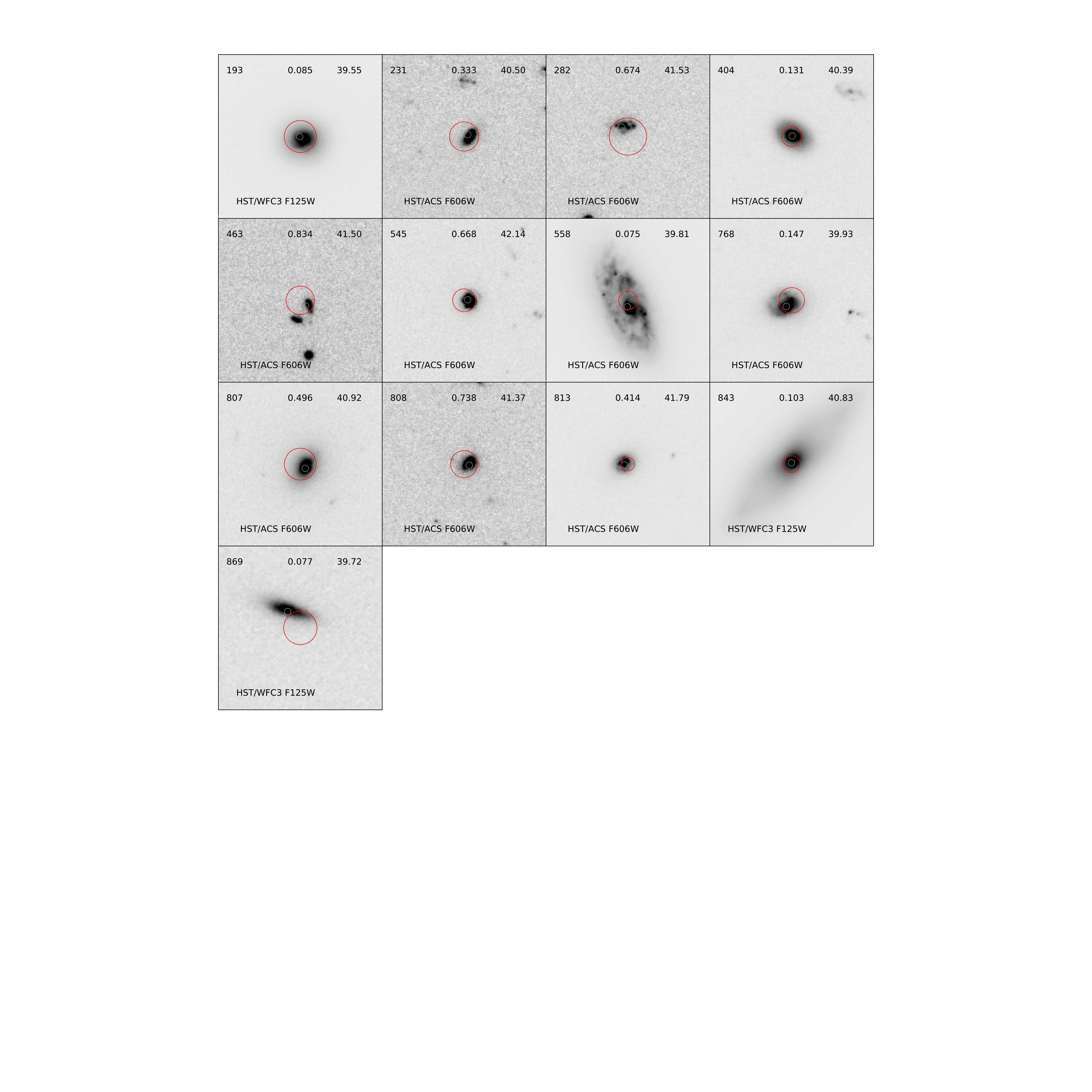}
\caption{Postage-stamp images in the GOODS-S/F606W (or CANDELS/F125W) band for the 13 variable sources. Each image is $4'' \times 4''$ with the position of the \mbox{X-ray} source of interest located at the center. The label at the top of each image gives XID, redshift, and logarithm of the intrinsic rest-frame 0.5--8~keV luminosity, respectively. The red and grey circles overplotted on each image are the positional uncertainties of the X-ray source and TENIS \mbox{$K_s$-band} counterpart at 90\% significance levels, respectively.}
\label{fig:Morphology}
\end{figure*}

We then compare the measured variability amplitude to that expected from XRB populations following an approach similar to that in Section~4.2 of \citet{2012ApJ...748..124Y}. To examine the potential contributions of XRB populations to variability, we first determine the relative contributions of HMXBs and LMXBs for each source. We calculate the expected \mbox{X-ray} luminosities from HMXBs ($L_{\mathrm{HMXB}}$) and LMXBs ($L_{\mathrm{LMXB}}$), respectively. Then, according to the ratio of $L_{\mathrm{LMXB}}/L_{\mathrm{HMXB}}$, the 13 variable sources are divided into two classes. 
One class is expected to have the \mbox{X-ray} luminosities dominated by HMXBs, and the other class is expected to have the \mbox{X-ray} luminosities dominated by LMXBs.
There are only two variable sources (XIDs 807, 843) whose expected \mbox{X-ray} luminosities are dominated by LMXBs. Based on the ``universal'' HMXB (LMXB) \mbox{X-ray} luminosity function, \citet{2004MNRAS.351.1365G} used Monte Carlo simulations to obtain a theoretical relation between $\sigma_{\mathrm{rms,tot}}/\sigma_{\mathrm{rms,0}}$ and SFR ($M_{\star}$) for the HMXB (LMXB) population, where $\sigma_{\mathrm{rms,tot}}$ is the square root of normalized excess variance of the whole HMXB (LMXB) population; $\sigma_{\mathrm{rms,0}}$ is the square root of normalized excess variance of an individual XRB, which can be as large as $20-30\%$ on $\sim$ year timescales \citep[e.g.,][]{2010LNP...794...17G}. We adopt $\sigma_{\mathrm{rms,0}} = 30\%$ to obtain the upper limit on $\sigma_{\mathrm{rms,tot}}^{2}$ as a function of SFR ($M_{\star}$) in the case of a HMXB (LMXB) dominant population.

Figure~\ref{fig:sigma2_SFR_M} displays SFR ($M_{\star}$) versus $\sigma^2_\mathrm{NXS, corr}$ for the \mbox{HMXBs-dominated} (LMXBs-dominated) variable sources (red stars), and the predicted upper limit on $\sigma_{\mathrm{rms,tot}}^{2}$ in the case of the HMXB (LMXB) dominant population is also shown (black solid curve). For reference, the data points of the 83 non-variable unclassified X-ray sources with available SFR and $M_{\star}$ measurements are also included in this figure (blue filled circles), where 76 non-variable unclassified \mbox{X-ray} sources with negative $\sigma^2_\mathrm{NXS, corr}$ values are placed at the $10^{-3}$ level and marked by the arrows.
The variability amplitude of the 13 variable sources are significantly higher than the upper limit of variability amplitude expected from XRB populations. This result indicates that the \mbox{X-ray} variability of the 13 variable sources are unlikely caused by XRB populations.

\subsection{The scenario of ULXs}
ULXs are another potential explanation for the \mbox{X-ray} variability of our selected variable sources. ULXs are defined as non-nuclear, point-like objects which at least once have been observed at an apparent isotropic X-ray luminosity higher than those of stellar-mass black holes \citep{2011NewAR..55..166F, 2017ARA&A..55..303K}. The typical \mbox{X-ray} luminosity range of ULXs is \mbox{$10^{39}-10^{41}$ erg~s$^{-1}$} \citep{2006AJ....131.2394L}. However, for the ULXs found in \mbox{early-type} galaxies, they have often lower luminosities, with $L_{\mathrm{0.5-8~keV, int}} < 2 \times 10^{39}$ erg s$^{-1}$ \citep{2004ApJS..154..519S}.

One method of identifying ULXs is to search for \mbox{off-nuclear} X-ray sources \citep[see][]{2004ApJ...600L.147H, 2006AJ....131.2394L, 2010A&A...514A..85M}. Figure~\ref{fig:Morphology} presents postage-stamp images ($4'' \times 4''$) of \emph{HST} F606W from \mbox{GOODS-S} \citep{2004ApJ...600L..93G} or F125W from CANDELS (\citealt{2011ApJS..197...35G}; \citealt{2011ApJS..197...36K}) for the 13 variable sources. The red and grey circles overplotted on each image are the positional uncertainties of the X-ray source ($\vartriangle_{X}$) and the TENIS $K_s$-band counterpart ($\vartriangle_{K_{s}}$\footnote{Following \citet[][]{2017ApJS..228....2L}, we adopt $0.1''$ as the $1\sigma$ positional uncertainty of TENIS \mbox{$K_s$-band} counterpart, so $\vartriangle_{K_{s}} \thickapprox 0.16''$ in here.}) at 90\% confidence levels, respectively. As in \citet{2006AJ....131.2394L}, we consider that an \mbox{X-ray} source is \mbox{off-nuclear} if its \mbox{X-ray} positional offset from the galactic nucleus is larger than the total positional uncertainty ($\vartriangle = \sqrt{\vartriangle_{X}^{2} + \vartriangle_{K_s}^{2}}$), where the position of the \mbox{TENIS} \mbox{$K_s$-band} counterpart is considered as the position of the galactic nucleus (the \mbox{$K_s$-band} data suffer less from dust obscuration or confusion from young \mbox{star-forming} regions). According to this criterion, there is one off-nuclear variable source XID~869. For this source, the offset angle between its \mbox{X-ray} position and its galactic nucleus is $1.1''$, and the confidence level of its \mbox{X-ray} position deviating from its galactic nucleus is $\approx98.8\%$. Since we searched for \mbox{off-nuclear} \mbox{X-ray} sources 13 times, we conservatively estimate that the probability of the deviation caused by chance is $\approx15.6\%$ (i.e., $13\times1.2\%$) after applying the Bonferroni correction\footnote{https://en.wikipedia.org/wiki/Bonferroni\_correction.}. Therefore, the positional offset is likely intrinsic, and XID~869 could be a ULX. It also has a low \mbox{X-ray} luminosity, with $L_{\mathrm{0.5-8~keV, int}} \approx 5 \times 10^{39}~\mathrm{erg~s}^{-1}$, which is consistent with the typical \mbox{X-ray} luminosities of ULXs.\footnote{The 0.5--8~keV intrinsic \mbox{X-ray} luminosity ($L_{\mathrm{0.5-8~keV, int}}$) is converted from $L_{\mathrm{0.5-7~keV, int}}$ based on the effective photon index.}

There are five variable sources (XIDs 282, 463, 545, 808, 813) whose luminosities are larger than the typical luminosities of ULXs (i.e., \mbox{$L_{\mathrm{0.5-8~keV, int}} > 10^{41}$ erg~s$^{-1}$}), which suggests they are likely not ULXs. There are seven variable sources whose luminosities ($L_{\mathrm{0.5-8~keV, int}}$) are less than \mbox{$10^{41}$ erg s$^{-1}$}. For four variable sources (XIDs 231, 404, 807, 843) whose luminosities are in the range of \mbox{$10^{40}-10^{41}$ erg s$^{-1}$}, their host galaxy morphologies are all \mbox{early-type} (see Figure~\ref{fig:Morphology}). Their higher X-ray luminosities (relative to \mbox{$2 \times 10^{39}$ erg s$^{-1}$}, the typical luminosity upper limit of ULXs found in \mbox{early-type} galaxies; e.g., \citealt{2004ApJS..154..519S}) suggest that they are likely not ULXs. For the remaining three variable sources (XIDs 193, 558, 768) in the luminosity range of \mbox{$10^{39}-10^{40}$ erg s$^{-1}$}, their variability due to a ULX near the galactic nucleus cannot be ruled out. Nevertheless, XID~558 and an unclassified \mbox{non-variable} \mbox{off-nuclear} \mbox{X-ray} source XID 556 are a pair of X-ray sources and have strong radio emission \citep[see][]{2017ApJS..228....2L}. They could be the radio core $+$ extended radio jet/lobe of a \mbox{radio-loud} AGN and are unlikely a pair of ULXs with radio emission. In addition, XID 558 also displays optical variability (see Section~3.3). Thus, XID~558 can be considered as a good AGN candidate.

\subsection{AGN candidates and their basic properties}
Through the above analyses, we find that for nine variable sources with luminosities ($L_{\mathrm{0.5-8~keV, int}}$) greater than \mbox{$10^{40}$ erg s$^{-1}$}, their \mbox{X-ray} variability is most likely attributed to the scenario of accreting SMBHs (i.e., AGNs). They are good AGN candidates. Among the four variable sources (XIDs 193, 558, 768, 869) with luminosities ($L_{\mathrm{0.5-8~keV, int}}$) less than \mbox{$10^{40}$ erg s$^{-1}$}, XID~869 is an off-nuclear source, which could be a ULX. 
XID 558 can be considered as a good AGN candidate (see Section~4.3). XID 193 and XID 768 are considered as AGN candidates in the following analyses, but we caution that the possibility that they are ULXs near the galactic nucleus cannot be excluded based on the current information. In summary, except for one variable source XID 869 which could be a ULX, we find 12 AGN candidates, of which 11 are LLAGN candidates ($L_{\mathrm{0.5-7keV, int}}<10^{42}$~erg~s$^{-1}$). Below we report the basic properties of the 12 AGN candidates. The remaining 97 \mbox{non-variable} unclassified X-ray sources and the ULX XID 869 are treated as normal galaxies in the following analyses.

The 12 AGN candidates all have optical spectroscopic observations \citep[][]{2004ApJS..155..271S,2005A&A...437..883M,2010A&A...512A..12B,2010ApJS..191..124S,2013A&A...549A..63K}, with spectroscopic redshifts ranging from 0.07 to 0.83. The optical spectra of the 12 AGN candidates reveal only narrow emission lines or absorption lines. There are four sources (XIDs 282, 404, 558, 813) which are classified by \citet{2004ApJS..155..271S} in more detail. XID~404 has a sufficiently large S/N in the optical spectrum for line-ratio measurement, and it is classified as a LINER through the line ratio diagnostics. The other three sources are classified as having unresolved emission lines consistent with an \mbox{HII} region-type spectrum.

\begin{figure*}[]
\centering
\includegraphics[width=6in]{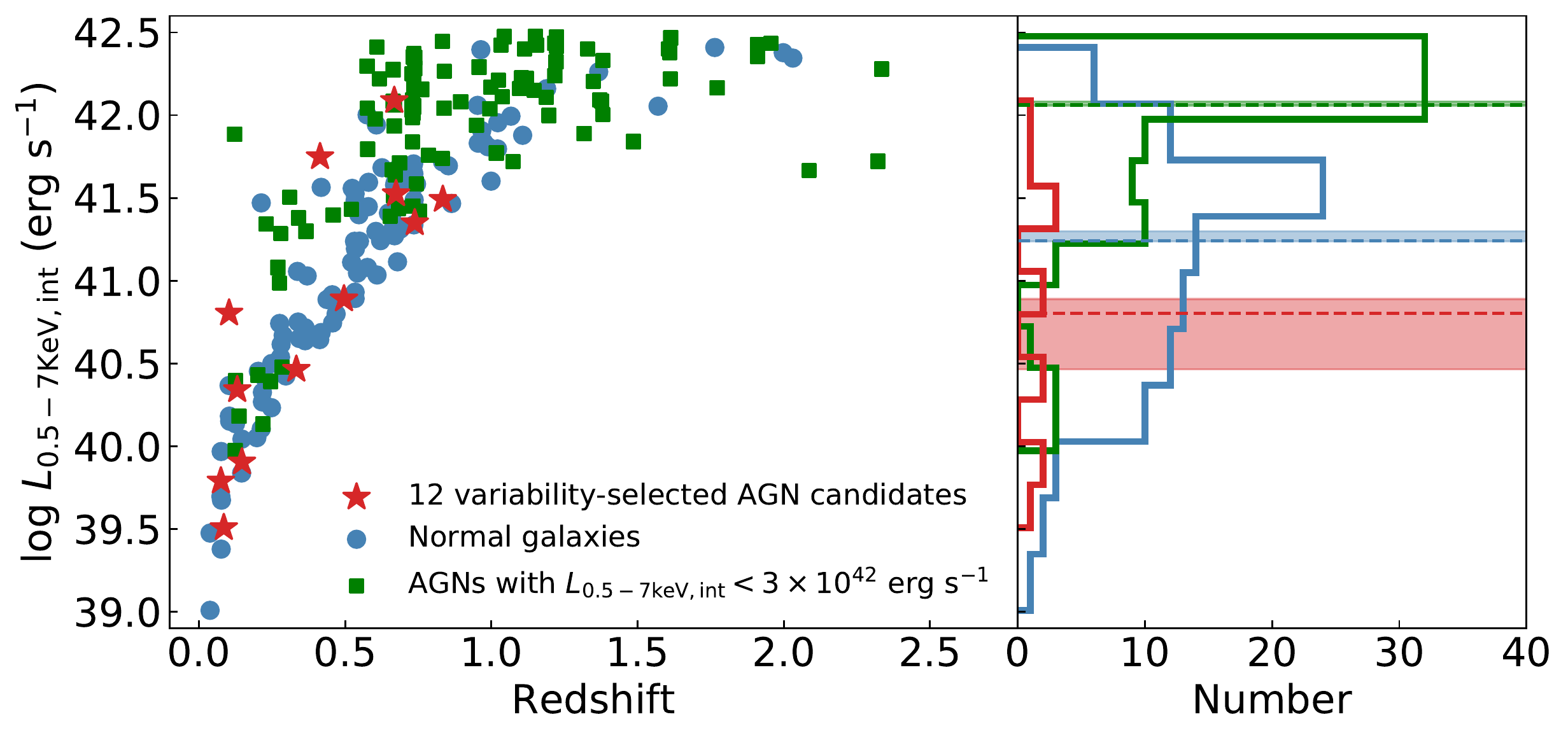}
\caption{Left panel: the intrinsic 0.5--7~keV luminosity vs. redshift distribution for the 12 variability-selected AGN candidates (red stars). The data points of the 98 normal galaxies (97 non-variable unclassified X-ray sources $+$ 1 ULX XID 869; blue filled circles) and 103 AGNs with \mbox{$L_{\mathrm{0.5-7keV, int}} <3 \times 10^{42}~\mathrm{erg~s}^{-1}$} (green filled squares) in our AGN sample are included for comparison. Right panel: the distribution of luminosities for the corresponding samples. The dashed lines are the median values of luminosities for the corresponding samples, and the shaded areas show $1\sigma$ uncertainties on the median luminosities, which are calculated from 10000 bootstrap samplings.
}
\label{fig:L_Z}
\end{figure*}

Figure~\ref{fig:L_Z} presents the intrinsic 0.5--7 keV luminosity versus redshift (left panel) and the luminosity distribution (right panel) for the 12 \mbox{variability-selected} AGN candidates. For comparison, the intrinsic \mbox{0.5--7}~keV luminosity versus redshift for the 98 normal galaxies (i.e., 97 non-variable unclassified \mbox{X-ray sources $+$ 1 ULX XID 869}) and 103 AGNs with \mbox{$L_{\mathrm{0.5-7~keV, int}} <3 \times 10^{42}~\mathrm{erg~s}^{-1}$} in our AGN sample (i.e., 395 AGNs) are also included. The 103 AGNs do not meet the luminosity criterion for selecting AGNs; they are selected by other criteria (see \citealt{2017ApJS..228....2L} for details). Of the 103 AGNs, only fourteen are classified as such using optical spectra. The median luminosity of the 12 variability-selected AGN candidates is significantly smaller than that of the 103 AGNs. A K-S test demonstrates that there is a significant difference between the luminosities of the 12 variability-selected AGN candidates and the 103 AGNs ($p = 8 \times 10^{-5}$). This result indicates that X-ray variability selection technique can effectively identify AGNs with low intrinsic \mbox{X-ray} luminosity (especially for $L_{\mathrm{0.5-7~keV, int}} <10^{41}~\mathrm{erg~s}^{-1}$) compared with other selection criteria.

The effective power-law photon indices ($\Gamma_\mathrm{eff}$) of the 12 AGN candidates were calculated in \citet{2017ApJS..228....2L} from the hard-to-soft band ratios by assuming that the \mbox{0.5--7~keV} spectra are power laws modified by only Galactic absorption. For the 12 AGN candidates, there are six sources which are only detected in the soft band or the hard band. For these sources, the best-guess values of band ratios were estimated by using the Bayesian code BEHR \citep{2006ApJ...652..610P}. The effective power-law photon indices of the 12 AGN candidates are listed in Table~\ref{Table2}, and the average effective power-law photon index is $1.82$. 

We also use XSPEC 12.9 \citep{1996ASPC..101...17A} to fit the \mbox{0.5--7 keV} spectra of the 12 AGN candidates. The \mbox{$C$-statistic} \citep{1979ApJ...228..939C} is employed to determine the best fit in the fittings. For five AGN candidates with numbers of net counts higher than 100, we fit their spectra individually. We use a \mbox{power-law} model modified by Galactic absorption (WABS*ZPOWERLW) to fit their spectra; an intrinsic absorption component is not required in all cases as it does not improve the fits. The fitting results for the five sources are listed in Table~\ref{Table5}. Their $\Gamma_\mathrm{XSPEC}$ values are in general consistent with the $\Gamma_\mathrm{eff}$ values given in \citet{2017ApJS..228....2L}. XID 545 has a hard photon index compared to ordinary AGNs, which may be caused by some intrinsic absorption that cannot be constrained by the current limited data. For the remaining seven sources, we use a power-law model modified by Galactic absorption with the same photon index to jointly fit their spectra. The best-fit photon index is $2.07^{+0.19}_{-0.18}$. This soft photon index indicates that the seven sources are, on average, not obscured.

\begin{table}
\tablenum{5}
\centering
\caption{\mbox{X-ray} spectral fitting results of five \mbox{variability-selected} AGN candidates with numbers of net counts higher than 100}
\begin{tabular}{@{}cccc@{}}
\hline\hline
XID	&	$\Gamma_\mathrm{XSPEC}$ &		Flux    \\
	&	 	&	($10^{-16}$ erg cm$^{-2}$ s$^{-1}$)	&	     \\
(1)	&	(2) 	&	(3)	\\
\hline
404	&	$1.51_{-0.19}^{+0.20}$	&			2.99  \\
545	&	$1.14_{-0.28}^{+0.29}$	&			4.60	\\
558	&	$1.95_{-0.34}^{+0.40}$	&			4.14  \\
813	&	$1.60_{-0.16}^{+0.16}$	&			6.8	\\
843	&	$1.98_{-0.13}^{+0.14}$	&			19.9	\\
\hline
\end{tabular}
\tablecomments{Column(1): Sequence number. Column(2): Best-fit power-law photon index obtained from spectral fitting and its 1$\sigma$ lower and upper uncertainties. Column(3): Galactic absorption-corrected flux in the full band (0.5--7 keV).}
\label{Table5}
\end{table}

Two (XID~193 and XID~558) of the 12 AGN candidates ($\sim 16\%$) are detected at 1.4~GHz by the VLA (see 7~Ms \mbox{CDF-S} main catalog in \citealt{2017ApJS..228....2L}). Their radio loudness parameters, $R_{\mathrm{L}} = f_{\nu}(\mathrm{5~GHz}) / f_{\nu}(\mathrm{4400~\AA})$, are calculated following \citet[][]{1989AJ.....98.1195K}. We convert $f_{\nu}(\mathrm{1.4~GHz})$ to $f_{\nu}(\mathrm{5~GHz})$ by assuming that the radio SED has a shape of $f_{\nu} \propto \nu^{-0.8}$. We estimate $f_{\nu}(\mathrm{4400~\AA})$ from the intrinsic \mbox{X-ray} luminosity of each source according to the average SED of AGNs; the observed optical flux is not adopted here because it is significantly contaminated by host galaxy starlight, which will severely bias the computed $R_{\mathrm{L}}$ value \citep[e.g.,][]{2017NatAs...1E.194P}. We adopt the average SED of type 1 quasars given by \citet[][]{2006ApJS..166..470R} and the average SED of AGNs with low accretion rates ($\lambda_{\mathrm{Edd}}<10^{-3}$) given by \citet[][]{2008ARA&A..46..475H} to estimate $f_{\nu}(\mathrm{4400~\AA})$. In the former case, the radio loudness parameters of XID 193 and XID 558 are $\sim100$ and $\sim218$, respectively. In the latter SED template, the radio loudness parameters are $\sim990$ and $\sim1800$, respectively. The results in both cases suggest that the two LLAGN candidates can be classified as radio-loud AGNs and may have strong relativistic jets.

\begin{figure}[]
\centering
\includegraphics[width=3.5in]{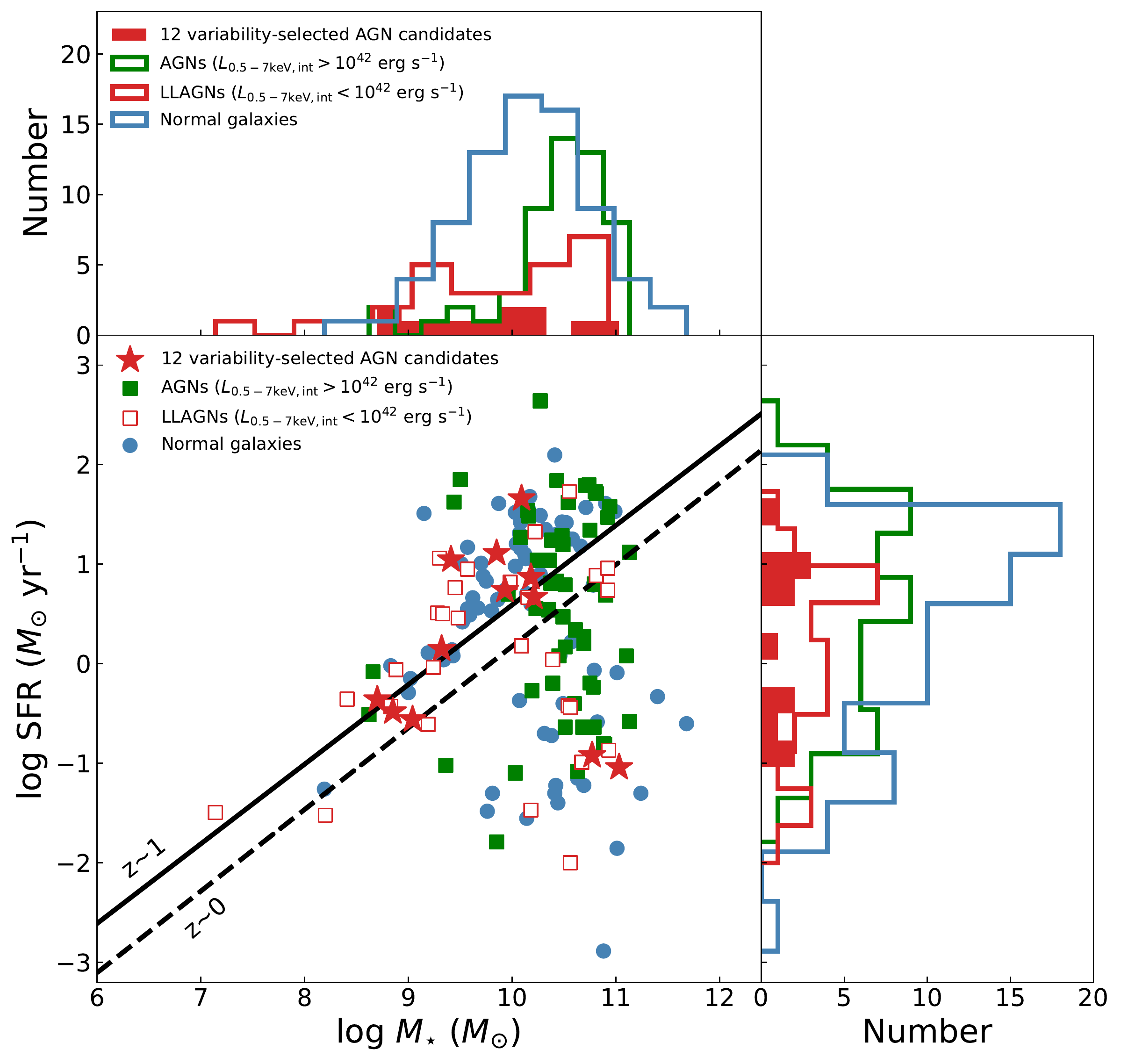}
\caption{Star formation rate (SFR) vs. stellar mass ($M_{\star}$) distribution for the 12 \mbox{variability-selected} AGN candidates (red stars). For comparison, the data points of 28 LLAGNs ($L_{\mathrm{0.5-7~keV, int}}<10^{42}~\mathrm{erg~s}^{-1}$; red open squares), 53 \mbox{higher-luminosity} AGNs ($L_{\mathrm{0.5-7~keV, int}}>10^{42}~\mathrm{erg~s}^{-1}$; green filled squares), and 75 normal galaxies (74 non-variable unclassified \mbox{X-ray sources $+$ 1 ULX XID 869}; blue filled circles), having $z<1$ and available SFR and $M_{\star}$ measurements, are plotted. The dashed line and solid line show the \mbox{star-forming} main sequences at $z\sim0$ and $z\sim1$, respectively \citep[][]{2007A&A...468...33E}.}
\label{fig:distribution}
\end{figure}

We also investigate the host-galaxy properties for the 12 \mbox{variability-selected} AGN candidates. The distribution of SFR versus $M_{\star}$ for the 12 host galaxies is shown in Figure~\ref{fig:distribution}. For comparison, the data points of 28 LLAGNs ($L_{\mathrm{0.5-7~keV, int}}<10^{42}~\mathrm{erg~s}^{-1}$) and 53 \mbox{higher-luminosity} AGNs ($L_{\mathrm{0.5-7~keV, int}}>10^{42}~\mathrm{erg~s}^{-1}$) in our AGN sample (i.e., 395 AGNs), and 75 normal galaxies (74 non-variable unclassified \mbox{X-ray sources $+$ 1 ULX XID 869}), having $z<1$ and available SFR and $M_{\star}$ measurements, are also included in this figure. The dashed line and solid line show the \mbox{star-forming} main sequences at $z\sim0$ and $z\sim1$, respectively \citep[][]{2007A&A...468...33E}. 
For the 12 AGN candidates ($0<z<1$), 10/12 are located above the \mbox{star-forming} main sequence at $z\sim0$ (7/12 above the \mbox{star-forming} main sequence at $z\sim1$), indicating that they tend to inhabit \mbox{star-forming} galaxies.
The \mbox{stellar-mass} distribution of their host galaxies is not significantly different from that of the 28 LLAGNs (K-S test $p=0.72$) selected by the other criteria in \citet{2017ApJS..228....2L} or the 75 normal galaxies ($p=0.22$), but it is significantly different from that of the 53 \mbox{higher-luminosity} AGNs ($p = 7\times10^{-4}$).
Most ($\sim72\%$) of the 53 \mbox{higher-luminosity} AGNs reside in high-mass galaxies ($\log M_{\star}>10.3$). For the 28 LLAGNs + 11 LLAGN candidates, $\sim31\%$ of the sources inhabit high-mass galaxies ($\log M_{\star}>10.3$), and $\sim44\%$ of the sources reside in \mbox{low-mass} dwarf galaxies ($\log M_{\star} < 9.5$).
The sources hosted in high-mass galaxies likely have black hole masses comparable to those of the \mbox{higher-luminosity} AGNs, and their low \mbox{X-ray} luminosities are probably attributed to inefficient accreting. Conversely, those sources inhabiting \mbox{low-mass} dwarf galaxies may be lower-mass black holes with typical accretion rates, similar to the low-luminosity quasars reported in \citet{2018MNRAS.480.1522L}.
In terms of the star formation rates of host galaxies, the distributions of the four samples (i.e., 12 AGN candidates, 28 LLAGNs, 53 \mbox{higher-luminosity} AGNs, and 75 normal galaxies) are not significantly different from each other.

\section{Discussion}
\subsection{The efficiency of X-ray variability selection of LLAGNs}
\begin{figure}[]
\centering
\includegraphics[width=3.7in]{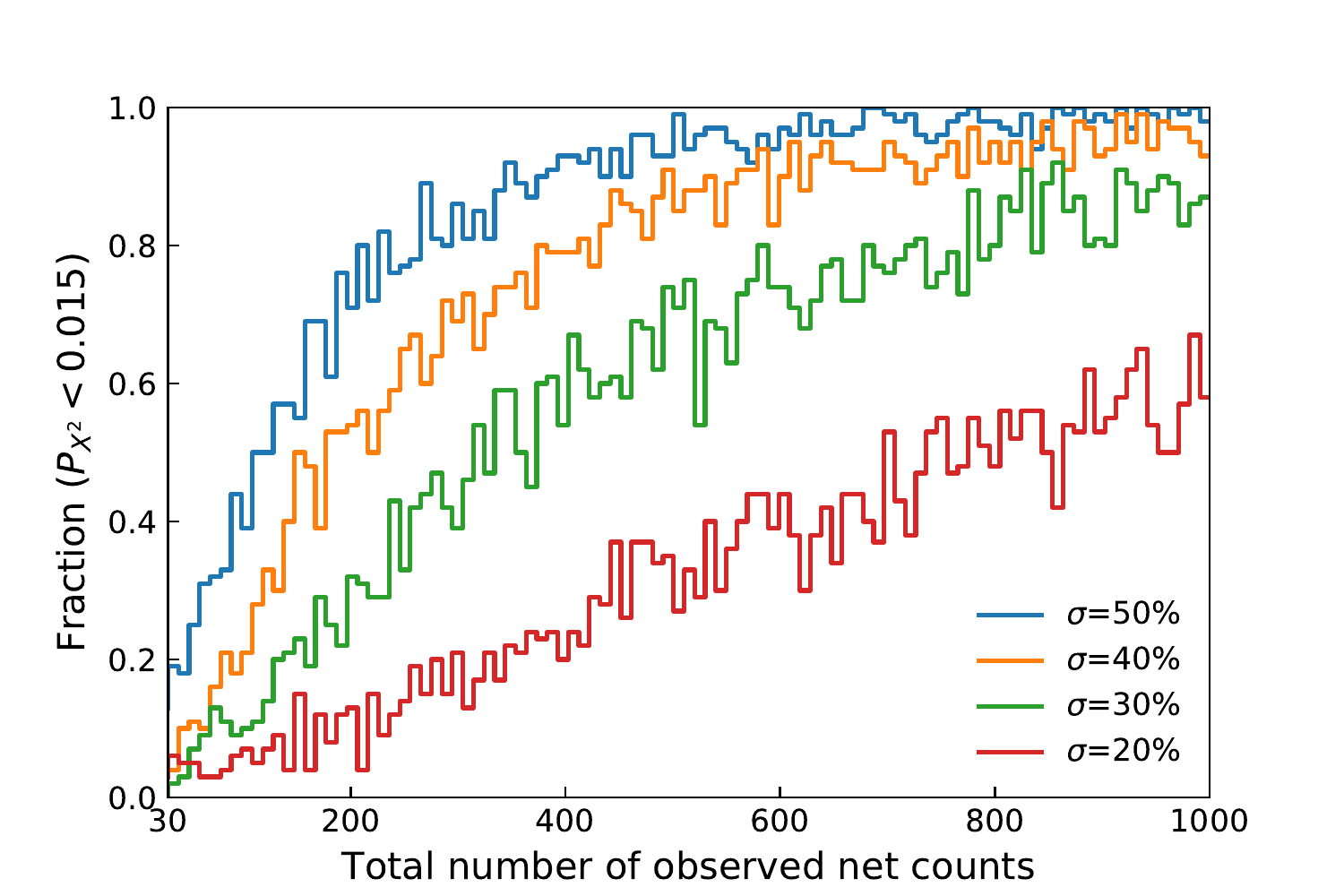}
\caption{The fractions of simulated light curves that have detectable variability vs. the total number of observed net counts. Four sets of fractions are presented under the assumption that AGN variability amplitude is $\sigma = 20\%$, $30\%$, $40\%$, and $50\%$, respectively.}
\label{fig:sim}
\end{figure}

As shown in Section~4.4, using the \mbox{X-ray} variability selection technique we can identify LLAGN candidates that are usually missed by other selection methods and have low intrinsic \mbox{X-ray} luminosity ($L_{\mathrm{0.5-7keV, int}} <10^{41}~\mathrm{erg~s}^{-1}$). However, the efficiency of \mbox{X-ray} variability selection of AGNs is critically dependent on the ability to detect \mbox{X-ray} variability. Generally, whether the variability of a source can be detected is related to the number of its observed net counts and its own variability amplitude given a fixed sampling pattern. We perform simulations to estimate the efficiency of detecting \mbox{X-ray} variable AGNs given different total numbers of observed net counts and variability amplitude. The specific procedure is described as follows:

We choose a source from 407 sources (12 variability-selected AGN candidates $+$ 395 AGNs) and then rescale the number of its observed net counts and variability amplitude (i.e., the variance of its photon flux fluctuation $\sigma^2$) to different values, where the number of observed net counts is rescaled by uniformly increasing the exposure time of each observation in each epoch, so that the photon fluxes of the source do not change. In each realization, we generate 1000 simulated light curves based on the method described in Section 4.1, and then calculate the $P_{X^2}$ value of each simulated light curve using the method described in Section 3.1. In this process, each simulated light curve is sampled based on the sparse observing pattern of the CDF-S survey and divided into seven epochs (i.e., Table~\ref{Table1}). Using $P_{X^2}<0.015$ as the threshold criterion for selecting variability, we obtain the fraction of simulated light curves that have detectable variability; the fraction reflects the ability to detect variability (i.e., the possibility that the variability of an AGN can be detected) under given conditions. We repeat this process for all 407 sources to obtain an average result.

The fractions as a function of the number of net counts are presented in Figure~\ref{fig:sim}, which displays four sets of fractions under the rescaling of the AGN variability amplitude to $\sigma = 20\%$, $30\%$, $40\%$, and $50\%$, respectively. As expected, the ability to detect variability increases with the increase of the total number of net counts and the variability amplitude of a source. When the total number of observed net counts is less than 100, even if the variability amplitude of the source is $\sigma = 50\%$ (above the average level of variability amplitude of AGNs, $\sim30-40\%$, e.g., \citealt{2004ApJ...611...93P, 2017MNRAS.471.4398P}), the probability that this level of variability can be detected is only $\sim 50\%$. This result strongly suggests that the 12 \mbox{variability-selected} AGN candidates are probably only the ``tip of the iceberg'' of unidentified LLAGNs, and some unclassified CDF-S X-ray sources may still host LLAGNs but do not exhibit detectable variability due to the small number of net counts measured.\footnote{The sparse observing pattern of the \mbox{CDF-S} survey and our binning strategy may also affect our ability to detect the variability of some unidentified LLAGNs with even considerable numbers of net counts.}
It is likely not feasible to identify variable LLAGNs by combining the sample objects into a group to increase the total number of net counts and searching collective variability, because a) individual LLAGNs in the group will likely not vary coordinately, which will reduce the amplitude of the collective variability and b) \mbox{non-variable} normal galaxies in the group will dilute the amplitude of the collective variability (see~Eq.~7).

The simulation results indicate that in the \mbox{low-count} regime, we tend to detect \mbox{X-ray} variability for sources with large variability. This \mbox{variability-detection} bias will lead to an overestimation of the measured overall variability amplitude of AGNs in the \mbox{low-count} regime due to the absence of a part of the AGNs that have weak and undetectable variability, as also suggested by previous studies \citep[e.g.,][]{2013ApJ...771....9A, 2017MNRAS.471.4398P}. 
In addition, the simulations demonstrate that when the total number of observed net counts is higher than 500 for AGNs with an average level of variability amplitude ($\sigma \approx30-40\%$), the probability of detecting their variability is $\approx60\%\textrm{--}80\%$.

In the above simulation, the $P_{X^2, \mathrm{4Ms}}$ and $P_{X^2, \mathrm{7Ms}}$ values for each simulated light curve can be calculated. Based on these two values for the 1000 simulated light curves, we can compute the proportion of the number of simulated light curves with $P_{X^2, \mathrm{4Ms}} < 0.05$ and $P_{X^2, \mathrm{7Ms}} > 0.05$ to the number of simulated light curves with $P_{X^2, \mathrm{4Ms}} < 0.05$. This proportion reflects the probability that the variability of a variable source cannot be detected when considering the 7~Ms data ($P_{X^2, \mathrm{7Ms}} > 0.05$) on the premise that its variability has been detected with the 4~Ms data ($P_{X^2, \mathrm{4Ms}} < 0.05$).
We calculate this probability for each of the eight variable sources mentioned in Section~3.3 and find that the probabilities range between $\approx \textrm{28--39}\%$.

\subsection{The fraction of variable AGNs }
\begin{figure}[]
\centering
\includegraphics[width=3.7in]{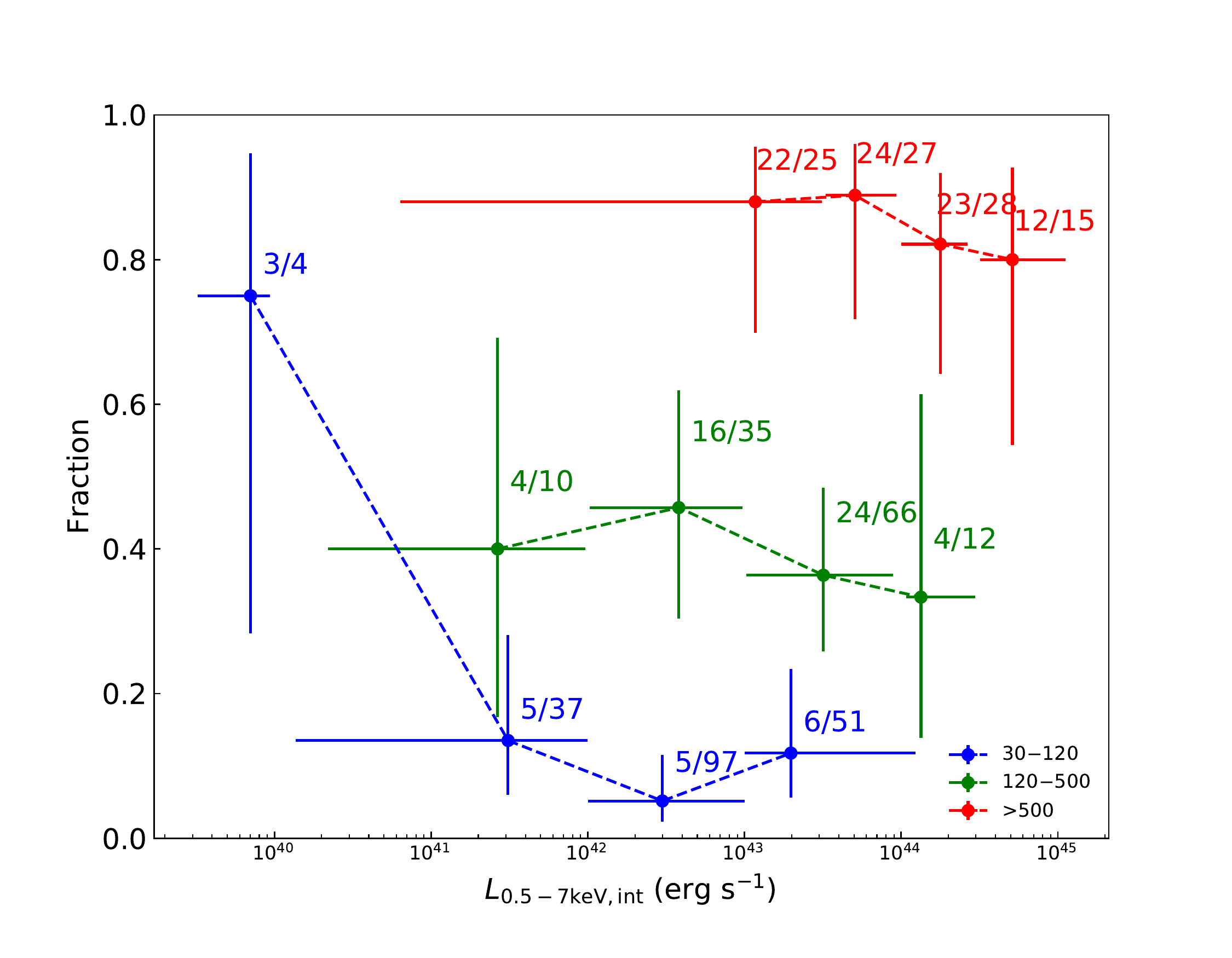}
\caption{The fractions of variable AGNs as a function of the intrinsic 0.5--7 keV luminosites in different net-count bins for the 407 AGNs (12 variability-selected AGN candidates $+$ 395 AGNs). The blue, green and red points represent fractions in the net-count bins of 30--120, 120--500, and $>500$, respectively. The horizontal error bars indicate the width of luminosity bin and the vertical error bars are the binomial errors on the fractions at a 95\% confidence level calculated following \citet{2011PASA...28..128C}. The labels are the ratio of the number of variable AGNs to the total number of AGNs in each bin.}
\label{fig:fraction}
\end{figure}

\begin{figure}[]
\centering
\includegraphics[width=3.7in]{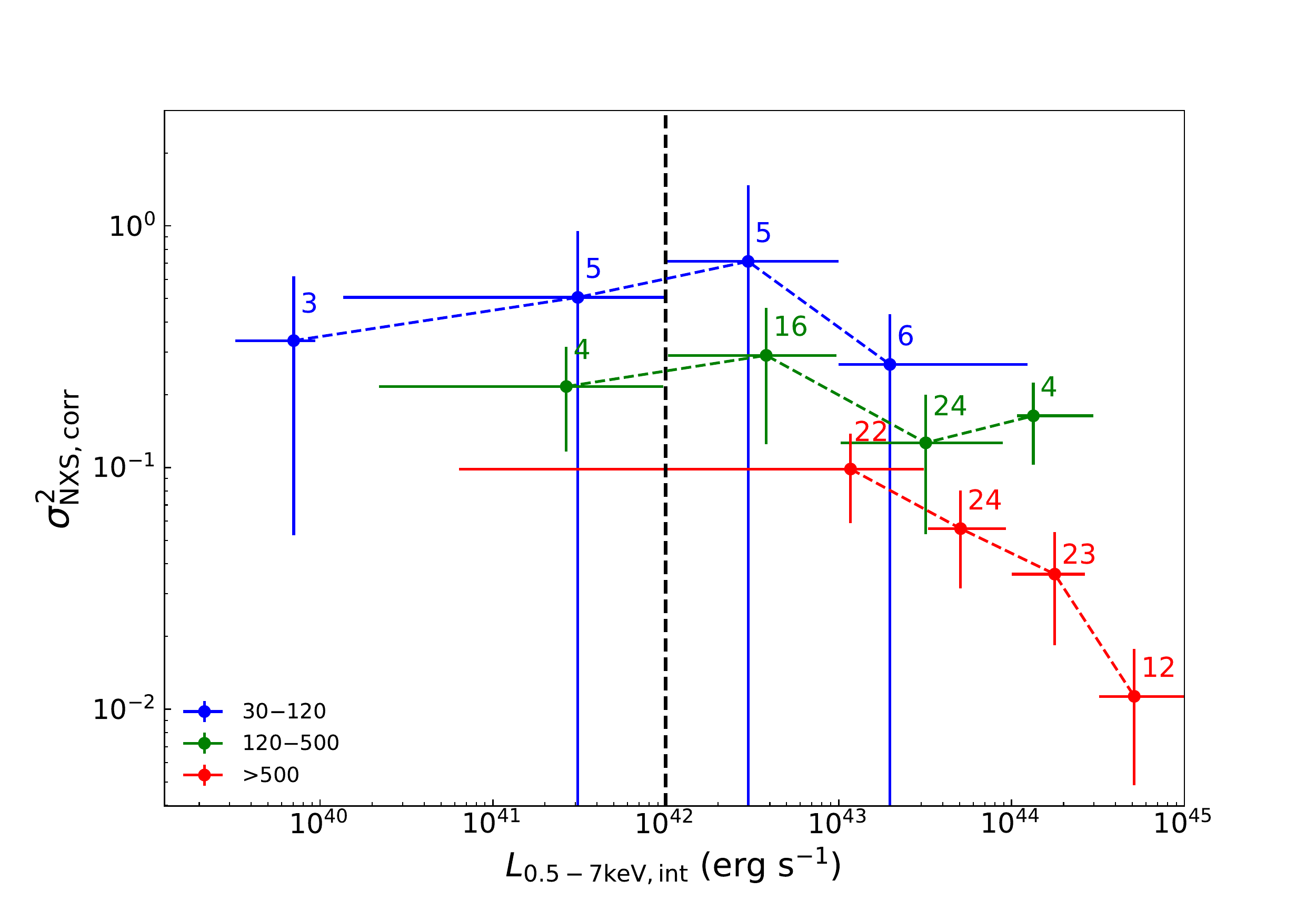}
\caption{The medians of  $\sigma^2_\mathrm{NXS, corr}$ vs. $L_{\mathrm{0.5-7keV, int}}$ in different net-count bins for the 148 variable AGNs (136 variable AGNs $+$ 12 variability-selected AGN candidates). The blue, green and red points represent the net-count bins of 30--120, 120--500, and $>500$, respectively. The horizontal error bars indicate the luminosity bin width, and the vertical error bars are the median of uncertainties on $\sigma^2_\mathrm{NXS, corr}$. The labels are the number of variable AGNs in each luminosity bin. The black dashed line is the typical LLAGN luminosity demarcation line ($L_{\mathrm{0.5-7keV, int}} \lesssim10^{42}$~erg~s$^{-1}$ for LLAGNs).}
\label{fig:sigma2_L}
\end{figure}

We investigate the fractions of variable AGNs in different luminosity bins. Using the method described in Section 3.1 ($P_{X^2}<0.015$ as threshold criterion for selecting variability), we find 136 variable AGNs from the 395 AGNs. Figure~\ref{fig:fraction} presents the fractions of variable AGNs as a function of their intrinsic 0.5--7 keV luminosities in different net-count bins taking into account the 12 variability-selected AGN candidates (i.e., the total number of sources is 395+12), where the size of each bin takes into account the need for dynamic changes in luminosity. Within the same net-count bins, the fractions of variable AGNs in bins with $L_{\mathrm{0.5-7keV, int}}  > 10^{40}~\mathrm{erg~s}^{-1}$ are the same within the errors. In the $>500$ net-count bin, the average fraction is~$\sim84.7\%$. In the 120--500 net-count bin, the average fraction is~$\sim38.9\%$. In the 30--120 net-count bin, the average fraction is~$\sim10.1\%$. These fractions are broadly consistent with the possibility that the variability of an AGN with typical variability amplitude ($\sigma \approx30-40\%$) can be detected given the corresponding number of net counts (see Figure~\ref{fig:sim}). This result demonstrates that X-ray variability is a near-ubiquitous property of AGNs with a wide range of luminosities, confirming the previous results \citep[e.g.,][]{2004ApJ...611...93P,2017MNRAS.471.4398P,2017arXiv171004358Z}. 

One abnormally high fraction is found in the 30--120 net-count bin. The fraction of variable AGNs in the luminosity bin with $L_{\mathrm{0.5-7keV, int}}  < 10^{40}~\mathrm{erg~s}^{-1}$ is higher than the fractions of variable AGNs in other luminosity bins at a $95\%$ confidence level. This significant discrepancy may be eased if we choose a different binning strategy for the luminosity bins. We also notice that there are two possible scenarios that can explain this abnormal fraction, described as follows:

1) This luminosity bin includes the variable objects XID~193, XID~558, and XID~768. As discussed in Section~4.3, the possibility that XID 193 and XID 768 are ULXs cannot be ruled out. Therefore this abnormal fraction may be due to the possibility that XID 193 and/or XID 768 are not AGNs but ULXs, leading to an overestimation of the fraction.

2) Within a given luminosity bin, a fraction of obscured AGNs may be missed by our sample selection due to their smaller numbers of observed net counts (after absorption) than our 30 net-count criterion. Such incompleteness is more pronounced in the low-count and low-luminosity regime. Therefore, in this low-luminosity bin, some obscured LLAGNs may be missed, leading to a significant underestimation of the denominator of the fraction of variable AGNs. In the same net-count bin (30--120 net-count bin), the fractions of variable AGNs for the other three \mbox{high-luminosity} bins are about the same within the errors and the average fraction is $\sim 10\%$. Assuming that the actual fraction of variable AGNs in this \mbox{low-luminosity} bin equals to this average value, we estimate that $\sim25$ obscured LLAGNs are missed in the sample selection in this \mbox{low-luminosity} bin.

\subsection{The luminosity-variability relation}
It has been established that \mbox{X-ray} variability amplitude is well anti-correlated with \mbox{X-ray} luminosity in high-luminosity AGNs for both short-term (less than a day, e.g., \citealt[][]{2012A&A...542A..83P, 2018arXiv180306891M}) and \mbox{long-term} (\mbox{month--year}, e.g., \citealt[][]{2004ApJ...611...93P, 2014ApJ...781..105L, 2016ApJ...831..145Y, 2017MNRAS.471.4398P, 2017arXiv171004358Z}) variability. However, the \mbox{luminosity-variability} relation in LLAGNs remains uncertain.

\cite{1998ApJ...501L..37P} studied a sample of LLAGNs observed with \emph{ASCA} on variability timescales of less than a day, and they found that the short-term variability amplitude of LLAGNs is significantly suppressed and does not follow the \mbox{luminosity-variability} relation of \mbox{Seyfert 1} galaxies. Subsequent studies found evidence for the short-term variability of LLAGNs being suppressed \citep[e.g.,][]{2004ApJ...606..173P, 2005ApJ...625L..39M}, but some other investigations found evidence arguing against this suppression \citep[e.g.,][]{2010MNRAS.401..677P, 2011A&A...530A.149Y}. The suppression of LLAGN short-term variability can be explained by changes in the accretion structure \citep[e.g.,][]{1998ApJ...501L..37P, 2004ApJ...606..173P}. It is generally believed that the accretion structure in LLAGNs differs from that in high-luminosity AGNs \citep[e.g.,][]{1996ApJ...462..142L, 2009MNRAS.399..349G, 2011A&A...530A.149Y, 2013A&A...556A..47H}. In LLAGNs, accretion rates are low; the accreted material cannot be effectively cooled to collapse into a standard thin disk, and the accretion flow is advection dominated \citep[e.g.,][]{2014ARA&A..52..529Y}. In this case, radiative cooling is inefficient, which causes the characteristic size of the X-ray emitting region to be larger than that in the standard accretion structure, resulting in the suppression of LLAGN short-timescale variability \citep[e.g.,][]{1998ApJ...501L..37P}.

\cite{2012ApJ...748..124Y} investigated the long-term variability of LLAGN candidates found from the 4~Ms \mbox{CDF-S} normal galaxies. Their results indicated that the long-term variability amplitude of LLAGN candidates with $L_{\mathrm{0.5-8keV, int}} <10^{41}~\mathrm{erg~s}^{-1}$ is below the extrapolated value of the \mbox{luminosity-variability} relation in high-luminosity AGNs and it is no longer inversely correlated to luminosity. However, because a few of their LLAGN candidates may be false detections, their results may be biased.

\begin{figure}[]
\centering
\includegraphics[width=3.6in]{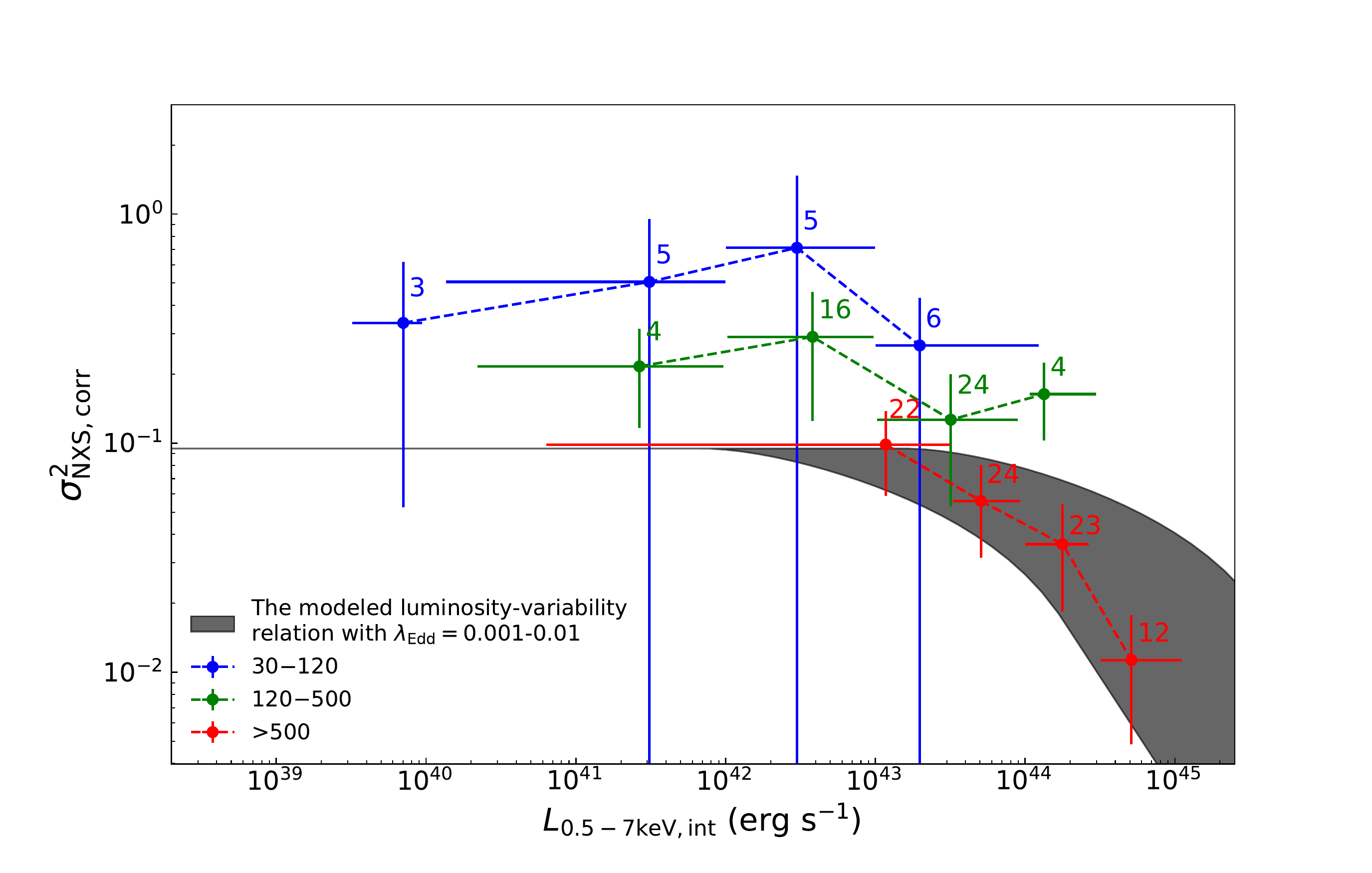}
\caption{The comparison of the observed data and the modeled luminosity-variability relation obtained from the empirical model with $\lambda_\mathrm{Edd}=0.01\textrm{--}0.001$. The observed data are the same as those in Figure~\ref{fig:sigma2_L}. The modeled luminosity-variability relation roughly reproduces the entire observed luminosity-variability trend.}
\label{fig:model}
\end{figure}

Below we investigate the luminosity-variability relation for \mbox{long-term} AGN variability. Figure~\ref{fig:sigma2_L} presents the medians of $\sigma^2_\mathrm{NXS, corr}$ versus $L_{\mathrm{0.5-7~keV, int}}$ in different net-count bins for the 148 variable AGNs ( 136 variable AGNs $+$ 12 \mbox{variability-selected} AGN candidates). In the high net-count bin ($>500$ counts bin), since the variability of most AGNs is detected, the medians of $\sigma^2_\mathrm{NXS, corr}$ unbiasedly reflect the overall variability amplitude of AGNs. In the low net-count bins (30--120 count bin and 120--150 count bin), there is a variability-detection bias (see Section~5.1) in the medians of $\sigma^2_\mathrm{NXS, corr}$, which leads to overestimations of the overall variability amplitude of AGNs in these bins. However, in each respective net-count bin, there is not a significant difference in the number of observed net counts for different luminosity bins, so we do not expect that such a bias would affect the \mbox{luminosity-variability} trend significantly in each respective net-count bin.
As shown in previous investigations of the CDF-S \citep[e.g.,][]{2012ApJ...748..124Y, 2016ApJ...831..145Y, 2017MNRAS.471.4398P, 2017arXiv171004358Z}, there is an anti-correlation between the $\sigma^2_\mathrm{NXS, corr}$ and $L_{\mathrm{0.5-7keV, int}}$ in the high-luminosity regime ($L_{\mathrm{0.5-7keV, int}} > 10^{42}$~erg~s$^{-1}$). However, with the decrease of luminosity, the variability amplitude of LLAGNs no longer follows the \mbox{anti-correlation} trend, and it appears independent of luminosity. This result is consistent with the finding of \cite{2012ApJ...748..124Y}.

The observed variability amplitudes of LLAGNs may be affected by the dilution of less variable \mbox{X-ray} emission from XRB populations. Using the values of $L_{\mathrm{2-10keV, XRB}}$ and $L_{\mathrm{2-10keV, int}}$ obtained in Section~4.2, we calculate the $L_{\mathrm{2-10keV, XRB}}$/$L_{\mathrm{2-10keV, int}}$ ratios of the 12 LLAGNs in the low-luminosity bins ($L_{\mathrm{0.5-7keV, int}}<10^{42}$~erg~s$^{-1}$) in Figure~\ref{fig:sigma2_L}. The lower and upper quartiles of the ratios are $\approx 6\%$ and $\approx 49\%$, respectively, which suggests that the observed variability amplitudes of some LLAGNs may be underestimated to some extent. Given the limited sample size and significant uncertainties associated with $\sigma^2_\mathrm{NXS, corr}$ and $L_{\mathrm{2-10keV, XRB}}$, it is not feasible to quantitively constrain this effect on the \mbox{luminosity-variability} trend in the low-luminosity regime.

\subsection{Interpreting the luminosity-variability relation with an empirical AGN variability model}
The \mbox{X-ray} variability of AGNs is usually described by a PSD function that is related to the accretion rate ($\lambda_\mathrm{Edd}$) and black hole mass ($M_\mathrm{BH}$) \citep[e.g.,][]{2004MNRAS.348..207P, 2006Natur.444..730M}. In this scenario, the \mbox{luminosity-variability} relation is the consequence of the dependence of the AGN PSD on the accretion rate and black hole mass \citep[see, e.g.,][]{2012ApJ...748..124Y, 2017MNRAS.471.4398P,2017arXiv171004358Z}. Below we show that the entire observed \mbox{luminosity-variability} trend can be roughly reproduced by an empirical AGN variability model based on a broken power-law PSD function that is applicable universally to our sample objects. The PSD is expressed as follows \citep[e.g.,][]{2011A&A...526A.132G}:

\begin{equation}
\label{eq16}
\mathrm{PSD}(\nu)=
\begin{cases}
C(\nu/\nu_{\mathrm{b}})^{-\beta}& (\nu \leq \nu_{\mathrm{b}})\\
C(\nu/\nu_{\mathrm{b}})^{-\alpha}& (\nu > \nu_{\mathrm{b}})\\
\end{cases}, 
\end{equation} 
where  $\nu_\mathrm{b}$ is the break frequency. \cite{2006Natur.444..730M} demonstrated that $\nu_\mathrm{b}$ depends on the accretion rate and black hole mass as $\nu_{\mathrm{b}} = 0.003\lambda_\mathrm{Edd}(M_\mathrm{BH}/10^{6}M_{\sun})^{-1}$. $C$ is the normalization of the PSD function. Following \cite{2004MNRAS.348..207P}, we assume that $C = 0.017/\nu_{\mathrm{b}}$. $\alpha$ and $\beta$ are the \mbox{high-frequency} and \mbox{low-frequency} spectral indexes, respectively. Previous studies of the \mbox{X-ray} variability of bright Seyfert galaxies found that $\alpha \approx 2$ and $\beta \approx 1$ \citep[e.g.,][]{2002MNRAS.332..231U, 2003MNRAS.341..496V, 2003MNRAS.339.1237V, 2009MNRAS.394..427B}.

Given a black hole mass and accretion rate, the theoretical excess variance of an AGN can be calculated based on the above empirical AGN variability model \citep[see, e.g.,][]{2017MNRAS.471.4398P, 2017arXiv171004358Z}. Assuming that the bolometric correction factors of AGNs depend on their accretion rates and follow Eq. 14 in \citet{2010A&A...512A..34L}, the relation between black hole mass and \mbox{X-ray} luminosity can be determined given an accretion rate, and then a modeled luminosity-variability relation for this given accretion rate can be derived (see, e.g., Section 5.4 of \citealt{2017arXiv171004358Z}). Figure~\ref{fig:model} presents the comparison of the observed data and the modeled \mbox{luminosity-variability} relation derived from the empirical model with $\lambda_\mathrm{Edd}=0.01\textrm{--}0.001$. The entire observed \mbox{luminosity-variability} trend, including an \mbox{anti-correlation} in the \mbox{high-luminosity} regime and a plateau in the \mbox{low-luminosity} regime, is roughly reproduced by the empirical model. In low net-count bins (30--120 count bin and 120--150 count bin), due to the variability-detection bias, the overall variability amplitude of AGNs is overestimated, so that the observed data points are higher than the modeled luminosity-variability relation. As mentioned earlier (Section~4.1), recent studies found that the \mbox{low-frequency} spectral index $\beta$ may not be 1 and suggested $\beta = \textrm{1.2--1.3}$ \citep[e.g.,][]{2017arXiv171004358Z}. We note that the $\beta$ value will affect the trend of the modeled luminosity-variability relation in the \mbox{low-luminosity} regime. However, due to the limited sample size in the \mbox{low-luminosity} regime and significant uncertainties associated with $\sigma^2_\mathrm{NXS, corr}$, it is not feasible to constrain the parameters of the PSD based on the current data.

\section{Summary and Future Prospects}
Using the \mbox{X-ray} variability selection technique, we search for LLAGNs that remain unidentified among the 7~Ms \mbox{CDF-S} \mbox{X-ray} sources. The main results are summarized in the following:

1. We find 13 variable sources from 110 unclassified \mbox{CDF-S} X-ray sources ($~12^{+4}_{-2}\%$) using a selection criterion of $P_{X^2}<0.015$, where $P_{X^2}$ is the probability that the observed photon flux fluctuation of a source is generated by Poisson noise alone.
See Section~3.

2. Except for XID 869 which could be a ULX, the variability of the remaining 12 sources is most likely attributed to accreting SMBHs (i.e., AGNs). These 12 variable sources are considered as AGN candidates, of which 11 are LLAGNs candidates. The redshifts of these AGN candidates range from 0.07 to 0.83. They are generally heavily obscured, with an average effective power-law photon index about 1.8. They have low intrinsic \mbox{X-ray} luminosities with a median luminosity of $7 \times10^{40}$~erg~s$^{-1}$ and tend to inhabit star-forming galaxies. 
Two of the 12 AGN candidates (XID 193 and XID 558) have large radio loudness and could be radio-loud AGNs. See Section~4.

3. Based on the simulation results of X-ray variability detection efficiency given different conditions, we find that when the total number of observed net counts is less than 100, even for an AGN with strong variability (above the average level of variability amplitude of AGNs), the probability of detecting its X-ray variability is only $\sim50\%$. This result suggests that the 12 \mbox{variability-selected} AGN candidates are probably only the ``tip of the iceberg'' of unidentified LLAGNs in the \mbox{CDF-S}. See Section~5.1. 

4. The fractions of variable AGNs are broadly consistent with the simulated X-ray variability detection efficiency. The fractions of variable AGNs are independent of X-ray luminosity and are only restricted by the number of observed net counts, confirming the previous findings that X-ray variability is a near-ubiquitous property of AGNs with a wide range of luminosities. See Section~5.2.

5. We confirm that there is an \mbox{anti-correlation} trend between X-ray luminosity and variability amplitude in high-luminosity AGNs ($L_{\mathrm{0.5-7keV, int}} > 10^{42}$~erg~s$^{-1}$) . However, with the decrease of the luminosity, the variability amplitude of LLAGNs ($L_{\mathrm{0.5-7keV, int}} < 10^{42}$~erg~s$^{-1}$) no longer follows the anti-correlation trend, and it appears independent of luminosity. An empirical AGN variability model based on a broken power-law PSD function can roughly reproduce the entire observed \mbox{luminosity-variability} trend, including an \mbox{anti-correlation} in the \mbox{high-luminosity} regime and a plateau in the \mbox{low-luminosity} regime. See Section 5.3 and 5.4.

In surveys by \mbox{next-generation} X-ray observatories, the \mbox{X-ray} variability selection technique can be used as an effective tool to select AGNs, complementing other AGN selection methods. For example, eROSITA \citep{2012arXiv1209.3114M} will repeatedly survey the entire sky in the \mbox{X-ray} band. With the \mbox{X-ray} variability selection technique, eROSITA is expected to detect $\sim60,000$ variable AGNs in the full sky \citep[][]{2017A&ARv..25....2P}. However, only a minor fraction ($\ll1\%$) of these AGNs could be LLAGNs according to the expected luminosity distribution of eROSITA-detected AGNs (Figure 5.2.3 in \citealt{2012arXiv1209.3114M}), and the reduced ability to detect the variability of low-luminosity/low-count AGNs (see Section~5.1). \emph{Athena} \citep{2013arXiv1306.2307N} has a large collecting area for detecting large numbers of \mbox{X-ray} photons, which makes it ideal for detecting the \mbox{X-ray} variability of LLAGNs. \emph{Athena} will be able to perform X-ray surveys more than two orders of magnitude faster than \emph{Chandra} \citep{2013arXiv1306.2307N} and thus will hopefully be able to produce a sample of LLAGNs that is 100 times larger than the current \mbox{CDF-S} sample in an exposure time comparable to the 7~MS \mbox{CDF-S}. With \mbox{X-ray} variability measurements of large samples of LLAGNs obtained from such surveys, we will be able to better understand the properties of LLAGN \mbox{X-ray} variability, the luminosity-variability relation in the \mbox{low-luminosity} regime, and the underlying PSD form.

\acknowledgments
We sincerely thank the anonymous referee for useful suggestions.
We thank Yong Shi and Dingrong Xiong for helpful discussions. We acknowledge financial support from the National Key R\&D Program of China grant 2016YFA0400702 (N.D., B.L.), National Natural Science Foundation of China grant 11673010 (N.D., B.L.), NASA ADP grant 80NSSC18K0878 (W.N.B.), National Key R\&D Program of China grant 2017YFA0402703 (Q.S.Gu), and National Natural Science Foundation of China grant 11733002 (Q.S.Gu). 
M.P. acknowledges support from the project "Quasars at high redshift: physics and Cosmology" financed by the ASI/INAF agreement 2017-14-H.0.
Y.Q.X. acknowledges support from the 973 Program (2015CB857004), NSFC-11473026, NSFC-11421303, and the CAS Frontier Science Key Research Program (QYZDJ-SSW-SLH006).

\end{document}